\documentclass[journal,twoside]{IEEEtran}
\usepackage{cite}
\usepackage{amsmath,amsfonts}
\usepackage{graphicx}
\usepackage{array}
\usepackage[caption=false,font=footnotesize]{subfig}
\usepackage{url}
\usepackage{color,soul}
\usepackage{multirow}
\usepackage[mdyy,hhmmss,12hr]{datetime}
\usepackage{booktabs}
\usepackage[colorlinks=true,allcolors=blue]{hyperref}
\usepackage{balance}

\graphicspath{{./fig/}{./fig/scatter/}}

\def\E{\mathbb{E}}
\def\P{\mathbb{P}}
\def\R{\mathbb{R}}
\def\e{{\bf e}}
\def\hatIt{\hat{\bf I}_t}
\def\I{{\bf I}}
\def\J{{\bf J}}
\def\r{{\bf r}}
\def\T{{\bf T}}
\def\H{{\bf H}}
\def\Q{\textrm{Q}}

\def\sQ{\mathcal{Q}}
\def\round{\operatorname{round}}
\def\mse{\operatorname{MSE}}
\def\var{\operatorname{Var}}
\def\pred{\text{pred}}
\def\psnr{\text{PSNR}}
\def\dequal{\stackrel{\text{def}}{=}}
\newcommand{\Hsup}[1]{H^{(#1)}}

\newcommand{\Rsup}[1]{R^{(#1)}}

\begin{document}
\title{Analysis of Coding Gain Due to In-Loop Reshaping}
\author{Chau-Wai~Wong,~\IEEEmembership{Member,~IEEE,}
Chang-Hong~Fu,~\IEEEmembership{Member,~IEEE,}
Mengting~Xu,
and~Guan-Ming~Su,~\IEEEmembership{Senior~Member,~IEEE}%
\thanks{This work was supported in part by China Scholarship Council under Grant 201806845028 and in part by the Key Research Development Plan of Jiangsu Province under Grant BE2023819. \textit{(Corresponding author: Chang-Hong Fu.})}%
\thanks{Chau-Wai Wong is with the Department of Electrical and Computer Engineering, NC State University, NC 27695 USA (e-mail: chauwai.wong@ncsu.edu).}%
\thanks{Chang-Hong Fu is with the School of Electronic and Optical Engineering, Nanjing University of Science and Technology, Jiangsu 210094, China (e-mail: enchfu@njust.edu.cn). He conducted the research work of this paper as a visiting scholar at the Dolby Laboratories.}%
\thanks{Mengting Xu conducted this work when she was with the School of Electronic and Optical Engineering, Nanjing University of Science and Technology, Jiangsu 210094, China.}%
\thanks{Guan-Ming Su is with the Dolby Laboratories, Sunnyvale, CA 94085 USA (e-mail: guanmingsu@ieee.org).}%
\thanks{Digital Object Identifier 10.1109/TIP.2024.3409189}
}

\IEEEpubid{1941-0042~\copyright~2024 IEEE. Personal use is permitted, but republication/redistribution requires IEEE permission.}

\maketitle

\begin{abstract}
Reshaping, a point operation that alters the characteristics of signals, has been shown capable of improving the compression ratio in video coding practices. Out-of-loop reshaping that directly modifies the input video signal was first adopted as the supplemental enhancement information~(SEI) for the HEVC/H.265 without the need to alter the core design of the video codec. VVC/H.266 further improves the coding efficiency by adopting in-loop reshaping that modifies the residual signal being processed in the hybrid coding loop. In this paper, we theoretically analyze the rate--distortion performance of the in-loop reshaping and use experiments to verify the theoretical result. We prove that the in-loop reshaping can improve coding efficiency when the entropy coder adopted in the coding pipeline is suboptimal, which is in line with the practical scenarios that video codecs operate in. We derive the PSNR gain in a closed form and show that the theoretically predicted gain is consistent with that measured from experiments using standard testing video sequences.
\end{abstract}


\section{Introduction}

The Joint Video Experts Team (JVET) started in 2017 the standardization process of Versatile Video Coding (VVC)/H.266~\cite{H266}, aiming at a 50\% bitrate reduction over the High Efficiency Video Coding (HEVC)/H.265~\cite{H265} while maintaining the subjective quality.
Among various newly introduced coding tools, the \textit{in-loop reshaping}\cite{xiu2018description} can improve the coding gain for both high dynamic range (HDR) \cite{lu2018hdrinloop} and standard dynamic range (SDR) \cite{lu2018sdrinloop, lu2018adaptive} video contents.

The use of the reshaping technique was initially motivated by the need to efficiently compress HDR video content using pre-HDR-era video codecs. 
Those video codecs were specifically designed and optimized for SDR videos and hence they do not perform well on HDR video content because of the mismatched characteristics between HDR and SDR videos.
SDR videos typically have a dynamic range of luminance from 0.1 nits to 100 nits \cite{dufaux2016high}.
It does not fully exploit the maximum range that the human visual system can simultaneously resolve, which is of four orders of magnitude~\cite{dufaux2016high}.
HDR videos, loosely defined as videos with a dynamic range greater than that of SDR videos and with an extended color space named the wide color gamut~(WCG), can significantly improve viewing experiences.
Due to the different sensitivity of human eyes at different luminance levels, HDR videos are usually represented in an alternate luminance domain, e.g., the Perceptual Quantization~(PQ) domain, rather than the Gamma domain traditionally used by SDR videos.
If one alters the signal characteristics of HDR video signal to mimic those of the SDR videos, video codecs such as H.264/AVC~\cite{H264} and HEVC/H.265 optimized for SDR videos can operate efficiently on HDR video contents.

\IEEEpubidadjcol

\begin{figure}[!t]
	\centering
		\subfloat[]{\includegraphics[width=0.5\textwidth]{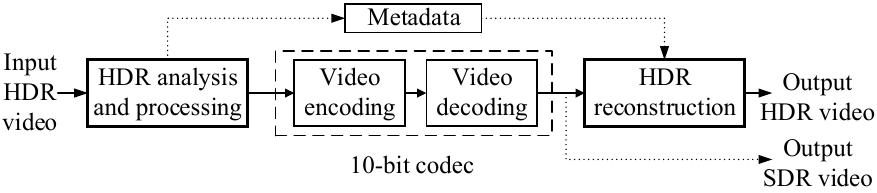}}\\
		\vspace{-3mm}\subfloat[]{\includegraphics[width=0.5\textwidth]{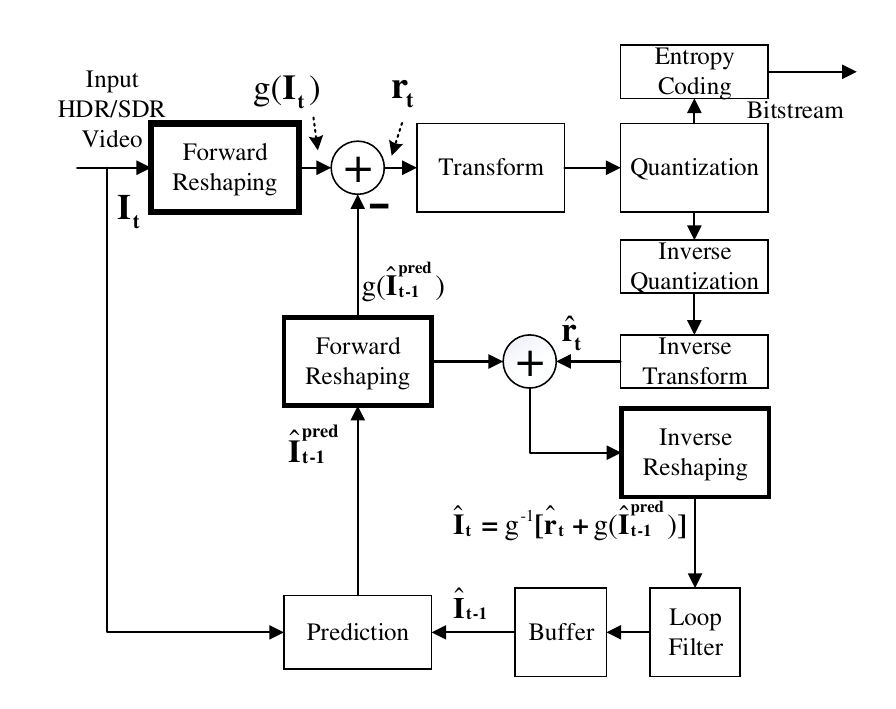}}\\
		\vspace{-3mm}\subfloat[]{\includegraphics[width=0.5\textwidth]{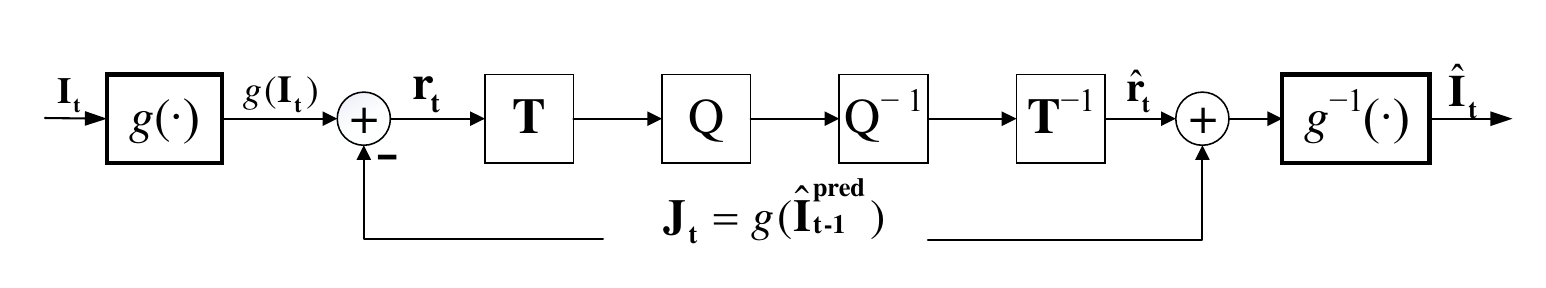}}
	\caption{Pipelines for video encoders with (a) out-of-loop reshapers and (b) in-loop reshapers. (c) Simplified schematic for the in-loop reshaper with blocks and symbols that are needed for the rate--distortion analysis.}
	\label{fig:pipeline}
\end{figure}

The first well-accepted technique to improve the coding efficiency of HDR videos was the \emph{out-of-loop} reshaping.
A Call for Evidence (CfE)\cite{cfe15083} was issued in February 2015 to evaluate the compression performance of the HEVC/H.265 on HDR/WCG videos and determine whether other promising technologies should be considered for future extensions of the standard.
The call was partly to address the wide emergence of HDR/WCG content after the HEVC/H.265 standard had been finalized.
Several proposals sharing similar ideas of using out-of-loop reshaper were submitted as responses~\cite{baylon2015response, baylon2015single, rusanovskyy2015single, ramasubramonian2015dual, lasserre2015technicolor, goris2015philips, cotton2015m36249, strom2015ericsson}, including approaches using a single layer and dual layers. 
In one joint response from Arris, Dolby, and Inter Digital~\cite{baylon2015response}, an adaptive intensity mapping was proposed as the pre- and postprocessing steps that encapsulate the HEVC/H.265 codec as illustrated in Fig.~\ref{fig:pipeline}(a). 
The characteristics of the input HDR video signal are first analyzed and then mapped to cope with the standard characteristics expected by the HEVC/H.265 Main 10 Profile encoder.
At the HEVC/H.265 decoder output, the signal is mapped backward to reconstruct the HDR signal for display.
As the pair of pre- and postprocessing mappings completely stay out of the hybrid video coding loops, such a strategy is named out-of-loop reshaping.
The change was proposed as an add-on to HEVC/H.265 because any major modification to then dominating standard would be costly.

As the standardization process for VVC/H.266 began, it became feasible to make major codec changes such as moving the reshaping operation into the predictive coding loop.
As time went on, several flaws in out-of-loop reshaping became evident to the video coding community.
First, the out-of-loop reshaping approach significantly constrained the design and optimization of the reshaper. 
The design and optimization could not be jointly performed, leading to marginal performance gain.
Second, the out-of-loop design led to the conformance point issue and end-to-end delay due to the extra postprocessing step.
Third, it also had to introduce additional on-chip buffers in the decoder, which is costly.
Because of the drawbacks and the opportunity from standardization, it was desirable to move the reshaper into the predictive coding loop for better coding gain and fewer issues.
Fig.~\ref{fig:pipeline}(b) illustrates an encoder with in-loop reshaping~\cite{lu2018hdrinloop} proposed by Dolby for VVC/H.266.
The forward and backward mappings are incorporated into the encoder design. In-loop reshaping was shown to be effective for not only HDR videos~\cite{lu2018hdrinloop,lu2018hdrinloop2} but also SDR videos~\cite{lu2018sdrinloop,lu2018sdrinloop2}. The in-loop reshaping tool was further optimized and renamed as \textit{luma mapping with chroma scaling (LMCS)}~\cite{lu2019mapping, lu2019simplification, lu2020LMCS}.

Although the in-loop reshaper had been
known in video coding practices to
improve the coding efficiency of HDR and SDR content~\cite{lu2018hdrinloop,lu2018sdrinloop,lu2018hdrinloop2,lu2018sdrinloop2} and was later adopted by VVC/H.266 version 1\cite{vvc} in July 2020, no theoretical justification has been provided by the video coding community.
Commonly asked questions include: (\textit{i}) Why can SDR video contents benefit from in-loop reshaping, while there is no such issue as a characteristic mismatch between the expected input and real input? 
Recall that the characteristic mismatch between the HDR contents and pre-HDR-era codecs was the initial motivation for adopting reshapers.
(\textit{ii}) Where does the coding gain come from and which factors do the gain depend on?
To address these questions, we will theoretically analyze a codec pipeline with an in-loop reshaper under a simplified reshaping scenario.
We will express in an analytic form the distortion and bitrate and investigate the rate--distortion change before and after reshaping.
We will also encode test video sequences using a simplified video codec to validate our theoretical model and result.
The key contribution of the paper is that, for the first time, the coding gain brought by the in-loop reshaping is theoretically justified.
Our analysis shows that such gain is independent of characteristics of the data (HDR or SDR, PQ or Gamma domain) and is a byproduct of the use of suboptimal entropy coders. 
We note that theoretical result holds even for 
the simplified reshaping scenario of using a one-piece reshaping function, i.e., a bounded range-expansion operation, instead of using a well-tuned multi-piece function adopted in VVC/H.266~\cite{vvc}. 
The analysis is amendable to multi-piece functions. 
We provide a preliminary analysis for the two-piece scenario and outline key steps for extending to multi-piece scenarios in Appendix~\ref{app:twopiece}.
Finally, we note that if real data and codec depart from the modeling assumptions adopted in this work, the actual coding gains could be different from the theoretical ones.

The rest of the paper is organized as follows. 
In Section~\ref{sec:def}, we define codec pipelines and symbols for theoretical analysis.
In Section~\ref{sec:theory}, we present the rate--distortion analysis with reshaping.
In Section~\ref{sec:exp}, we use testing video sequences to verify the theoretical result.
In Section~\ref{sec:discuss}, we discuss some common concerns.
In Section~\ref{sec:conclude}, we conclude the paper.

\section{Definitions of a Codec with In-Loop Reshaper}
\label{sec:def}

\begin{figure}[!t]
  \centering
  \includegraphics[width=1.0\linewidth]{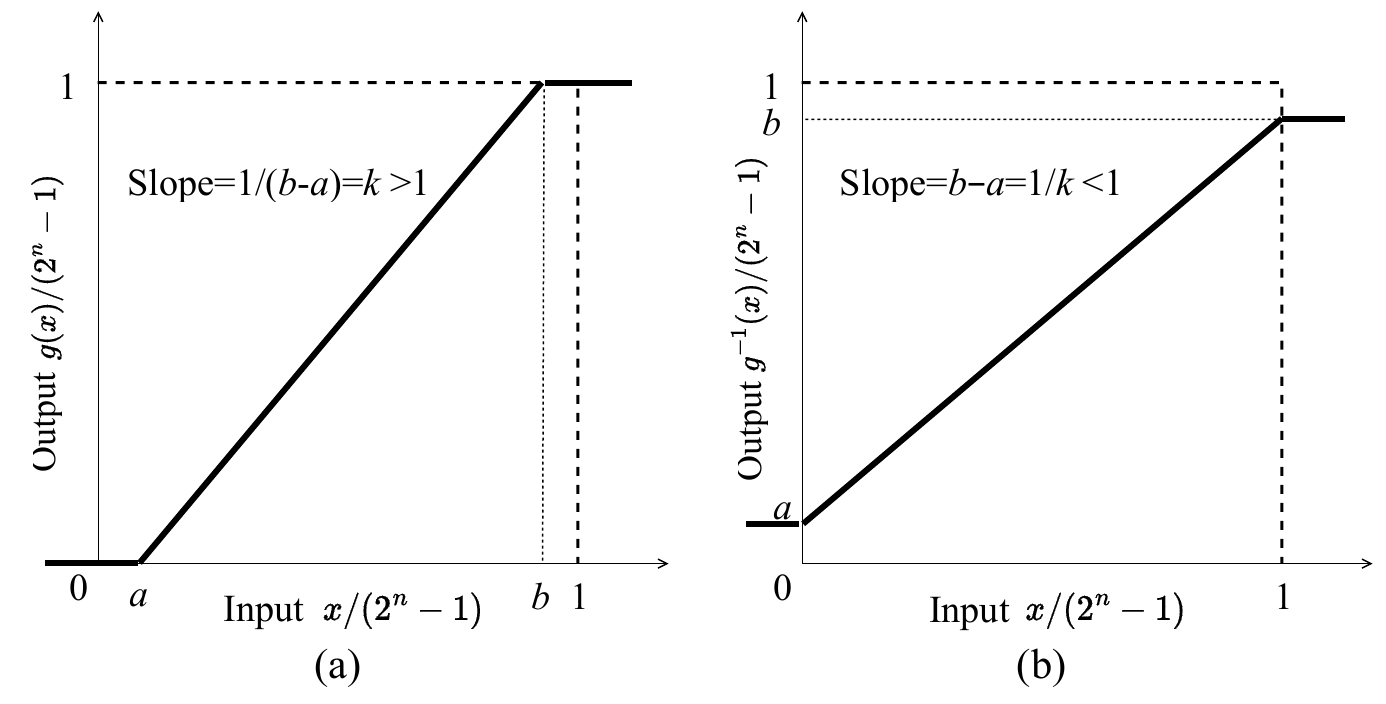}
  \caption{A typical one-piece (a) forward reshaping function $g(x) \in [0, M]$, where $M = 2^n-1$, and (b) its corresponding backward reshaping function $g^{-1}(x) \in [aM, bM]$, where $0 < a < b < 1$.}
  \label{fig:reshape_map}
\end{figure}

Fig.~\ref{fig:pipeline}(b) illustrates how in-loop reshaping blocks, i.e., those highlighted with thicker borders, may be incorporated into a standard hybrid video coding encoder.
We follow the decoder definition in \cite[Fig.~4]{lu2018adaptive} and extrapolated a complete encoder schematic accordingly. 
Here, the forward reshaping function (or the forward reshaper) is a monotonically nondecreasing function $g(x)$ that acts elementwise on a raw or predicted input video signal to maximize its range. 
Reshaping is more commonly known as point processing in the image processing literature~\cite[Ch.~3.1.1]{gonzales2008digital}.
Fig.~\ref{fig:reshape_map} shows an example of the forward reshaping function $g$, in which the range $[aM, bM]$ of an input signal will be linearly mapped to the range of $[0, M]$, where $0 < a < b < 1$, $M = 2^n - 1$, and $n$ is the number of bits representing the intensity of a pixel.
For example, to expand the range of 10-bit video input to the maximum possible range,
the output of $g(x)$ should range from $0$ to $1023$.
Conversely, the backward reshaping function (or the backward reshaper) $g^{-1}(x)$ compresses the full-range reconstructed video signal back to the range of the raw video signal.

As illustrated in Fig.~\ref{fig:pipeline}(b), the forward reshaper maps the raw frame $\I_t$ and predicted frame $\hat{\I}_{t-1}^\pred$ into a reshaped domain before the residual frame $\r_t$ is calculated. 
The transform coding, quantization, and frame reconstruction all operate 
in the reshaped domain. 
The backward reshaper maps the reconstructed frame back to the nonreshaped domain for deblocking, buffering, display, and motion compensation. 
In practical schemes, $g$ is a nonlinear, oftentimes implemented by a multipiece linear function~\cite{bross2021overview, H266}.
We will see from both the theoretical analysis and experiments of this paper that even if the nonlinear mapping has only one piece as in Fig.~\ref{fig:reshape_map}, there will be a gain in coding efficiency. 

We simplify Fig.~\ref{fig:pipeline}(b) by removing processes/blocks that are irrelevant to the rate--distortion analysis.
An equivalent diagram with symbols similarly defined as in~\cite{wong2016impact} is shown in Fig.~\ref{fig:pipeline}(c). 
$\I_t$ and $\hatIt \in [0, 2^n-1]^N$ 
are the raw frame to be encoded and the reconstructed frame at time $t$ in the nonreshaped domain, where $N$ is the total number of pixels in a frame.
$g(\cdot)$ is the deterministic forward reshaping function. 
$\hat{\I}_{t-1}^{\pred}$ is the motion compensated frame in the nonreshaped domain. The raw residual frame $\r_t$ is generated by subtracting the reshaped motion-compensated frame $\J_t \stackrel{\text{def}}{=} g(\hat{\I}_{t-1}^{\pred})$ from the reshaped raw frame $g(\I_t)$. 
$\T$ is the forward transform. 
$\Q(\cdot)$ and $\Q^{-1}(\cdot)$ are the elementwise quantization and the dequantization operations. 
We will use their composition, namely, the quantizer $\sQ(\cdot) = \Q^{-1}[\Q(\cdot)]$, for the following analysis. 
$\hat{\r}_t$ is the reconstructed residual frame in the reshaped domain.

To facilitate matrix--vector analysis of the encoder pipeline, 
we consider $\I_t$, $\hatIt$, $\r_t$, and $\hat{\r}_t$ column vectors of length $N$, and $\T \in \R^{N \times N}$ a combined transform matrix.
To accommodate the use of a block-based 2-d separable transform, e.g., a $4$-by-$4$ discrete cosine transform (DCT), we arrange the pixels of vector $\r_t$ in groups following the scanning order of transform blocks. Within each group, all pixels are vectorized columnwise.
The corresponding horizontal transform $\H_4$ and vertical transform $\H_4^T$ applied to a 2-d block can be represented by applying the transform matrix $\T_{16} = \H_4 \otimes \H_4^T$~\cite{jain1989fundamentals} to the vector created from the 2-d block, where $\otimes$ is the Kronecker product.
Putting the transforms for all blocks together, we can construct a sparse, orthogonal, block-diagonal combined matrix $\T$ by repeating $\T_{16}$ on the diagonal.

\section{Rate--Distortion Analysis of with Reshaper}
\label{sec:theory}
We prove in this section that range expansion leads to coding gain for suboptimal entropy coders. We start by deriving the expressions of distortion, entropy, and bitrate as functions of the quantization step $q$ and the range expansion slope $k$. We obtain rate--distortion functions by canceling $q$ and reveal the analytic expression of coding gain.

\subsection{Distortion Reduction Due to Range Expansion}
We first show the impact of in-loop reshaping on the distortion.
We write the relations among symbols of the equivalent diagram in Fig.~\ref{fig:pipeline}(c) from left to right as follows:
\begin{subequations}
\begin{align}
\r_{t} &= g(\I_t) - g\big(\hat{\I}_{t-1}^{\pred}\big), \label{eq:pipeline_eq1} \\
\hat{\r}_{t} &= \T^{-1} \sQ \left(\T \r_{t}\right), \label{eq:pipeline_eq2} \\
\hatIt &= g^{-1}\left(\hat{\r}_{t} + g \big(\hat{\I}_{t-1}^{\pred} \big) \right). \label{eq:pipeline_eq3}
\end{align}
\end{subequations}
Substituting the reconstructed residual frame (\ref{eq:pipeline_eq2}) into (\ref{eq:pipeline_eq3}), we obtain:
\begin{subequations}
\begin{align}
\hatIt &= g^{-1}\Big(\T^{-1} \sQ \left(\T \r_t
\right) + g\big(\hat{\I}_{t-1}^{\pred}\big) \Big) \label{eq:pipeline_eq4} \\
&= g^{-1} \Big(\T^{-1} \big[ \T\r_t + q \cdot e\big( \T\r_t / q \big)\big] + g\big(\hat{\I}_{t-1}^{\pred}\big) \Big) \label{eq:pipeline_eq5} \\
&= g^{-1} \big( g(\I_t) + \T^{-1} q \cdot e\big( \T\r_t / q \big) \big), \label{eq:pipeline_eq6}
\end{align}
\end{subequations}
where (\ref{eq:pipeline_eq6}) can be obtained using (\ref{eq:pipeline_eq1}), and (\ref{eq:pipeline_eq5}) can be obtained by using the definition of the quantizer that decomposes the quantized result $\sQ(x)$ into the sum of the raw input $x$ and a rounding residue as follows:
\begin{equation}
\sQ(x) = q \cdot \round(x/q)
= x + q \cdot e (x/q),
\end{equation}
where $q \in \mathbb{R}^+$ is the quantization step, $\round(\cdot)$ is the rounding to the nearest integer operation, and the rounding residue $e(x)$ is defined as follows:
\begin{equation}
e(x) = \round(x) - x, \quad e(x) \in\left(-1/2, 1/2\right].
\end{equation}
We focus on the simplest case that there is only one nonflat linear segment in the reshaping function as shown in Fig.~\ref{fig:reshape_map},
namely, 
\begin{subequations}
\begin{align}
g(x) &= 
\begin{cases}
0, & x \le aM,\\
k(x - aM), & aM < x \le bM, \\
M, & x > bM,
\end{cases} \label{eq:forward-reshape}\\
g^{-1}(x) &= 
\begin{cases}
a, & x \le 0,\\
x/k + aM, & 0 < x \le M, \\
b, & x > M,
\end{cases}
\end{align}
\end{subequations}%
where $0 < a < b < 1$ and $k = 1 / (b-a) > 1$ for range expansion.
We note that only the middle branches of the forward reshaper $g(x)$ and the backward reshaper $g^{-1}(x)$ are inverse to each other. The full functions are not inverse to each other due to the clipping branches in their definitions. With these two points in mind,
we further simplify (\ref{eq:pipeline_eq6}) as follows:
\begin{align}
\hatIt 
\approx \I_{t} - k^{-1} \T^{-1} q \cdot e \big( \T\r_t / q \big),
\label{eq:full_pipeline_eq_final}%
\end{align}%
where the approximation is caused by the coordinates of the input to $g^{-1}$ in~\eqref{eq:pipeline_eq6} whose values are less than $0$ or greater than $M$.
We provide rigorous proof in Apppendix~\ref{app:derivation_recon_err} that these cases are of low likelihood and do not contribute significantly to the overall reconstruction error.

Rearranging~\eqref{eq:full_pipeline_eq_final}, we calculate the reconstructed squared error for frame $\I_t$ as follows:
\begin{subequations}
\begin{align}
\left\|\hatIt - \I_{t}\right\|^{2}
&\approx \left\|k^{-1} \T^{-1} q \cdot e\Big( \T\r_t / q\Big)\right\|^{2} \\
&= (q/k)^{2}\left\| e \Big( \T\r_t / q \Big) \right\|^{2}. \label{eq:raw_mse_final}
\end{align}
\end{subequations}
Note that $\T^{-1}$ disappears from the expression as it is an orthogonal matrix that does not affect the norm of any vector it multiplies to. The rounding residue function $e(\cdot)$ is applied elementwise to each pixel of the $(1/q)$-scaled transformed residual frame $\T\r_t / q$, resulting $N$ nearly uncorrelated random variables uniformly distributed from $-1/2$ to $1/2$.
We define $\mse(\hat{\I}_{t}, \I_{t})$ as the expected mean squared error averaged over all pixel locations in a frame, namely,
\begin{equation}
\mse(\hat{\I}_{t}, \I_{t}) = \E \left[ \frac{1}{N} \left\|\hat{\I}_{t} - \I_{t}\right\|^{2} \right]
\approx k^{-2} \cdot \frac{q^2}{12}.
\label{eq:mse}
\end{equation}
The above result reveals that when there is no reshaping, i.e., $k = 1$, the MSE expression degenerates to $q^2/12$, a well-known result for a uniform quantizer being used in high-rate scenarios~\cite{widrow2008quantization}.
Our result in (\ref{eq:mse}) shows that the in-loop reshaper is effective in reducing the MSE by a multiplicative factor of $k^{-2} < 1$ in the range expansion scenarios.

\subsection{Entropy Increase Due to Range Expansion}
Next, we find the entropy change due to reshaping using the method of differential entropy given its analytic tractability.
Except for discrete uniform random variables, it is difficult to obtain an analytic expression for the entropy of a quantized random variable even if the random variable's PDF has a parametric form.
The theory of differential entropy (reviewed in Appendix~\ref{sec:didff-entropy-background}) suggests that the entropy change before and after reshaping roughly equals the change in differential entropy.
To measure differential entropy, we focus only on the transformed residues before quantization. Without loss of generality, we denote one particular coordinate of transformed residues before and after reshaping by $X$ and $Y$, respectively.
Since the residues are zero mean, range expansion using \eqref{eq:forward-reshape} on the input signal $\I_t$ will result in a linear relation%
\footnote{We note that $\T \big(g(\I_t) - g(\hat{\I}_{t-1}^{\pred})\big) = k \cdot \T \big(\I_t - \hat{\I}_{t-1}^{\pred} \big)$ for most of time except in rare cases that $\hat{\I}_{t-1}^{\pred} \notin [0, M]$. We ignore the rare cases as their proportions in entropy are low.} 
between the transformed residue, namely, $Y = kX$. 
We skip detailed proof and state that the range expansion due to reshaping increases the entropy by roughly $\log_2 k$, namely,
\begin{equation}
H^{(1)} \approx H^{(0)}+\log_2 k,
\label{eq:H1_additive}
\end{equation}
where $H_0$ and $H_1$ are the entropy of the quantized residues before and after reshaping, respectively.
The detailed proof can be found in Appendix~\ref{sec-app:proof-range-expan-boost-entro}.

\subsection{No Gain Using Optimal Entropy Coder}
\label{subsec:no_gain}
\begin{figure}[!t]
  \centering
  \subfloat[]{\includegraphics[clip, trim=0.5mm 0 25 13, width=0.485\linewidth]{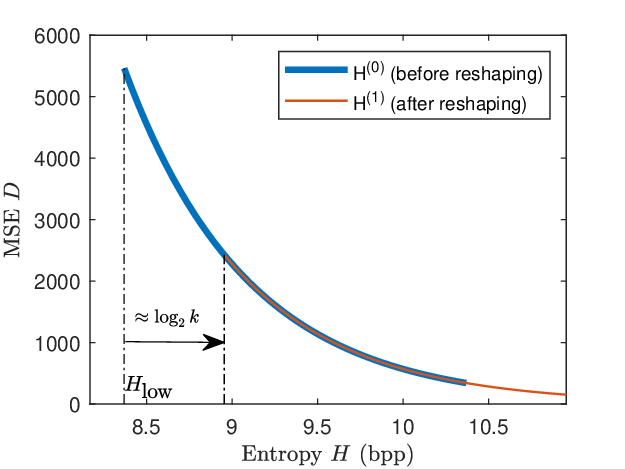}}
  \hspace{1mm}
  \subfloat[]{\includegraphics[clip, trim=0.5mm 0 25 13, width=0.485\linewidth]{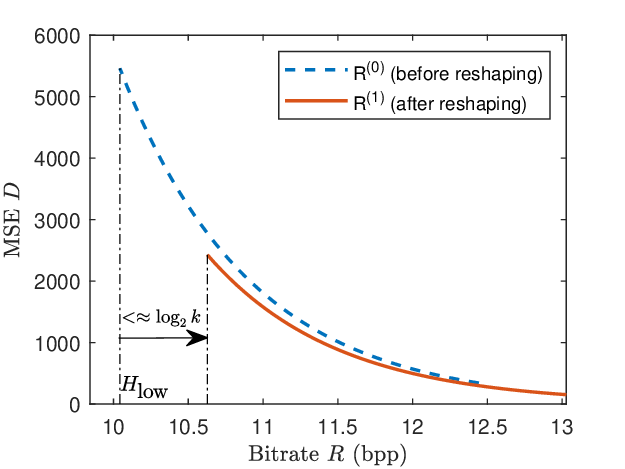}}
  \caption{(a) MSE--entropy~(H--D) curves overlap before and after reshaping, indicating that reshaping does not lead to coding gain for optimal entropy coders. 
  (b)~MSE--bitrate~(R--D) curves before and after reshaping with a range expansion factor $k=1.5$ for a suboptimal entropy coder with a slope ratio $\eta=0.83$. The gap indicates that reshaping can lead to coding gain for a suboptimal entropy coder.}
  \label{fig:mse_vs_H}
\end{figure}

We now show that in-loop reshaping does not lead to any coding gain in the rate--distortion sense if the entropy coder is optimal.
We assume the uniform distribution for the residual frame,%
\footnote{The residual signal distributes closer to Gaussian due to the central limit theorem. We use uniform distribution for illustration purposes. One can easily show that the PSNR gain is the same for Gaussian distributed residues.}
$H^{(0)}=-\sum p_{i} \log_2 p_{i} =\log_2 (2^n/q)$, where $2^n/q$ is the number of codewords for $n$-bit videos.
Combining with \eqref{eq:H1_additive}, we obtain 
\begin{equation}
H^{(1)} \approx \log_2 \left(\frac{2^n}{q / k}\right). \label{eq:H1_uniform}
\end{equation}
Recall in the MSE derivation in (\ref{eq:mse}), we have shown that $\mse = (q/k)^2/12$, 
hence the (entropy, distortion) points before and after reshaping when assuming the uniform distribution can be written as:
\begin{subequations}
\begin{align}
(H^{(0)}, D^{(0)}) &= \left(\log_2 \left(\frac{2^n}{q}\right), \frac{q^{2}}{12}\right),\\
(H^{(1)}, D^{(1)}) &= \left(\log_2 \left(\frac{2^n}{q / k}\right), \frac{(q / k)^{2}}{12}\right).
\end{align}
\end{subequations}
It is easy to show that for both cases, the H--D curves have the following expression:
\begin{equation}
D = \frac{1}{12} \cdot 2^{-2(H-n)},
\end{equation}
except that they have shifted supports $\mathcal{H}^{(0)} = [H_\text{low}, H_\text{high}]$ and $\mathcal{H}^{(1)} = \mathcal{H}^{(0)} + \log_2 k$, respectively. Here, $H_\text{low}$ and $ H_\text{high}$ are the lower and upper limits for the interval $\mathcal{H}^{(0)}$, respectively.
Fig.~\ref{fig:mse_vs_H}(a) illustrates a pair of H--D curves before reshaping~(blue) and after reshaping~(red).
Hence, we conclude that if an optimal entropy coder is used, i.e., rate equals entropy, reshaping will not bring gain in the rate--distortion sense given that the operating point merely moves along the same H--D curve.

\subsection{Gain Due to Suboptimal Entropy Coder}
\label{subsec:gain-suboptimal}
\begin{figure}[!t]
  \centering
  \includegraphics[width=0.9\linewidth]{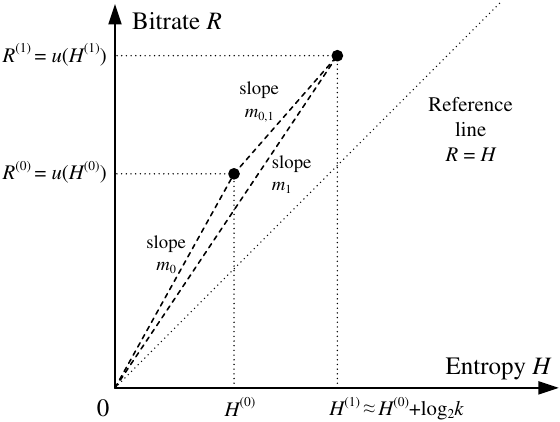}
  \caption{Operating points $(H^{(0)}, R^{(0)})$ and $(H^{(1)}, R^{(1)})$ on bitrate function $u(\cdot)$ (not shown for clarity) before and after reshaping on the bitrate--entropy plane, respectively. The secant from the origin to $(H^{(0)}, R^{(0)})$ has a slope of $m_0 > 1$ and the secant from $(H^{(0)}, R^{(0)})$ to $(H^{(1)}, R^{(1)})$ has a slope of $1 \leq m_{0,1} < m_1$. The very fact that the slope ratio $\eta = m_{0,1} / m_0 < 1$ ensures that reshaping can always lead to coding gain when suboptimal entropy coders are used.}
  \label{fig:br_vs_H}
\end{figure}

We now extend the entropy result from the previous subsection to bitrate and show that in-loop reshaping can lead to coding gains for suboptimal entropy coders.
Entropy coders in video codecs are almost always suboptimal. Mathematically, we define a \textit{bitrate function} $u(\cdot)$ of a practical entropy coder relating the bitrate $R$ and the entropy $H$, which must satisfy the following relation:
\begin{equation}
R = u(H) > H.
\label{eq:suboptimal_mapping}
\end{equation}
In the bitrate--entropy~(R--H) plane of Fig.~\ref{fig:br_vs_H}, all possible $u(\cdot)$ (not shown to reduce the complexity of the figure) will result in operating points laying in the upper half region above the line of $R = H$.
For a given video codec with bitrate function $u(\cdot)$, we use points $(\Hsup{0}, \Rsup{0})$ and $(\Hsup{1}, \Rsup{1})$ to denote the operating conditions before and after reshaping, respectively:
\begin{subequations}
\begin{align}
R^{(0)} &= u \left(H^{(0)}\right) \label{eq:def0}, \\
R^{(1)} &= u \left(H^{(1)}\right). \label{eq:def1}
\end{align}
\end{subequations}
Using \eqref{eq:def0} and \eqref{eq:suboptimal_mapping}, it is immediate that the slope $m_0$ of the secant line for $u(\cdot)$ between $(0, 0)$ and $(H^{(0)}, R^{(0)})$ satisfy the following relation:
\begin{align}
  m_0 \stackrel{\text{def}}{=} R^{(0)} / H^{(0)} \in (1, \infty).
  \label{eq:slope-m0}
\end{align}
Similarly, we define the slope $m_1$ of the secant line for $u(\cdot)$ between $(0, 0)$ and $(H^{(1)}, R^{(1)})$ as
\begin{align}
m_1 \dequal R^{(1)} / H^{(1)} \in (1, m_0).
\label{eq:slope-m1}
\end{align}
Here, we assume that $m_1$ is upper bounded by $m_0$ (see Fig.~\ref{tab:exp_rd}(c) as an example), i.e., the entropy coder operates more efficiently at a higher rate $(H^{(1)}, R^{(1)})$ after range expansion than at $(H^{(0)}, R^{(0)})$.
As our final theoretical result in \eqref{eq:psnr_gain_theory} will show, when $m_1 \to m_0$, the PSNR gain vanishes no matter how large the range is expanded.
We also define the slope $m_{0,1}$ of the secant line for $u(\cdot)$ between $(\Hsup{0}, \Rsup{0})$ and $(\Hsup{1}, \Rsup{1})$ as 
\begin{equation}
  m_{0,1} \dequal \frac{\Rsup{1} - \Rsup{0}}{\Hsup{1} - \Hsup{0}} \in [1, m_1).
  \label{eq:m01-def}
\end{equation}
Here, the upper bound of $m_{0,1} < m_1$ is due to the geometry of the triangle formed by the origin and the two operating points. 
The lower bound of $m_{0,1} \geq 1$ is reasonable since the increase in bitrate should be at least comparable to the increase in entropy.

Given the definitions of the slopes of the three secant lines, we can continue to simplify the relation after reshaping described in \eqref{eq:def1} as follows:
\begin{subequations}
\begin{align}
R^{(1)} 
&= u\left(H^{(1)}\right) \\
&\approx u\left(H^{(0)}+\log_2 k\right) \label{eq:R1-expand-H0H1relation} \\ 
&\approx u\left(H^{(0)}\right) + m_{0,1} \cdot \log_2 k \label{eq:taylor}\\
&= m_0 \cdot H^{(0)} + (\eta \cdot m_0) \log_2 k \label{eq:r1_decomposition}\\
&\approx m_0 \cdot \big( h^{(0)} - \log_2 q \big) + (\eta \cdot m_0) \log_2 k \label{eq:R1-dif-entropy-approx}\\
&= \log_2 \bigg(\frac{2^{n + h^{(0)}}}{q/k^{\eta}} \bigg)^{m_0}, \label{eq:R1-expand-final-form}
\end{align}
\end{subequations}
where \eqref{eq:R1-expand-H0H1relation} is due to the information-theoretic approximation~\eqref{eq:H1_additive}, \eqref{eq:taylor} is due to the geometry, and \eqref{eq:R1-dif-entropy-approx} is due to the approximation of discrete entropy by differential entropy in \eqref{eq:H_h_relation}.
In \eqref{eq:r1_decomposition}, we introduce the \textit{slope ratio} $\eta$ defined as
\begin{align}
\eta &= m_{0,1} \big/ m_0 \in (0,1),
\label{eq:def-eta}
\end{align}
where the upper bound is due to $m_{0,1} < m_1 < m_0$.
The slope ratio $\eta$ may partially reflect the optimality of the entropy coder.
For example, for an optimal entropy coder, all operating points should be on $R = H$, hence $\eta = 1$.
As we will see later, the slope ratio $\eta$ is one of the only two factors that determine the PSNR gain.
Reorganizing~\eqref{eq:R1-expand-final-form}, we obtain the rate--distortion operating points before and after reshaping for a nonideal entropy coder as follows:
\begin{subequations}
\begin{align}
(R^{(0)}, D^{(0)}) &= \left(\log_2 \bigg(\frac{2^{n + h^{(0)}}}{q}\bigg)^{m_0}, \frac{q^{2}}{12}\right),\\
(R^{(1)}, D^{(1)}) &= \left(\log_2 \bigg(\frac{2^{n + h^{(0)}}}{q/k^{\eta}} \bigg)^{m_0}, \frac{(q / k)^{2}}{12}\right).
\end{align}
\label{eq:subopt-theory-rd}%
\end{subequations}
It is easy to show via canceling the quantizer step $q$ that the R--D curves have different expressions before and after expansions as follows:
\begin{subequations}
\begin{align}
D &= \frac{1}{12} \cdot 2^{-2 \left[ R/m_0 - (n + h^{(0)}) \right]},\\
D &= \frac{1}{12} \cdot 2^{-2 \left[ R/m_0 - (n + h^{(0)}) \right]} \cdot k^{2(\eta-1)}.
\end{align}
\label{eq:DR-theory}%
\end{subequations}
Hence, the gain in PSNR given the same bitrate $R$ is
\begin{equation}
\Delta \psnr = 10 \log_{10} \big( D^{(0)} / D^{(1)} \big) = 20(1-\eta) \log_{10} k.
\label{eq:psnr_gain_theory}
\end{equation}
Note that the PSNR gain is dependent on the slope ratio $\eta$ and the reshaping scalar $k$, but independent of its differential entropy $h^{(0)}$, i.e., the shape of the histogram of the residual signal before reshaping.
The PSNR gain is also independent of the bitrate $R$ specific to the current operating condition hence theoretically, the gain is uniform over $R$.
TABLE~\ref{tab:psnr_gain_theory} illustrates a few calculated examples of the PSNR gain under typical values for $\eta$ and $k$.
\begin{table}[!t]
\renewcommand{\arraystretch}{1.3}
\caption{Theoretically Calculated PSNR Gain\vspace{-2mm}}
\label{tab:psnr_gain_theory}
\centering
\begin{tabular}{ccc}
\toprule
$\eta$ & $k$ & $\Delta$PSNR (dB) \\ \hline
0.95 & 2.0 & 0.30 \\
0.98 & 2.0 & 0.12 \\ \hline
0.95 & 1.8 & 0.26 \\
0.98 & 1.8 & 0.10 \\ \hline
1 & Any value & 0 
\\ \bottomrule
\end{tabular}
\end{table}
Fig.~\ref{fig:mse_vs_H}(b) illustrates a pair of simulated R--D curves with range expansion factor $k = 1.5$ when the suboptimality of the entropy coder is configured to have $m_0 = 1.2$ and $m_{0,1} = 1$. The R--D curve after reshaping has a better rate--distortion trade-off as predicted by our theory.

\section{Experimental Results on Test Sequences} \label{sec:exp}
\subsection{Use of Simplified Video Codec}

To verify the theoretical PSNR gain due to in-loop reshaping as predicted by our theoretical result in~\eqref{eq:psnr_gain_theory}, we implemented a simplified video codec with 
an IPP$\dots$ group-of-pictures structure with the macroblock size of $16\times16$ and the DCT size of $4\times4$.
Since any practical video codec is suboptimal, our simplified codec can allow us to calculate the two entropy--bitrate points before and after reshaping as illustrated in Fig.~\ref{fig:br_vs_H}. 
An off-the-shelf codec such as VVC/H.266 was not used because its sophisticated structure makes it difficult to calculate the corresponding entropy and bitrate.
The entropy coder in our implementation is customized to allow tuning the degree of suboptimality of the coder, which allows us to comprehensively verify the theoretical result~\eqref{eq:psnr_gain_theory} from the previous section. 

The block diagram of the codec is shown in Fig.~\ref{fig:pipeline}(b). Motion estimation with full search and motion compensation is performed for each macroblock in P frames. 
Full search can avoid potential inaccuracies caused by a fast motion estimation algorithm.
Search ranges are determined per test sequence based on the statistics of motion vectors extracted from VVC/H.266 compressed files and the computational complexity of the full search. 
More details on search range determination are given in Appendix~\ref{app:search_range} of the supplemental material.
Residual blocks are DCT-transformed and quantized. 
Arithmetic coding is then performed on the quantized DCT coefficients to obtain the bitrate. Meanwhile, entropy is calculated based on the histogram of codewords/quantized coefficients. In order to approximate the adaptive behavior of context-based adaptive binary arithmetic coding (CABAC) used in standardized video codecs, we 
construct the histogram of codewords at the frame level. The three reshaping blocks with thick borders in Fig.~\ref{fig:pipeline}(b) are bypassed if reshaping is turned off.

\subsection{Experimental Conditions}
Using the simplified video codec, we were able to verify the proposed theoretical model on standard test sequences 
\textsc{BasketballDrill}, 
\textsc{BQMall}, 
\textsc{FourPeople}, 
\textsc{KristenAndSara}, 
\textsc{PartyScene}, and 
\textsc{Vidyo}.
Summaries of visual activities in the test sequences are provided in Appendix~
\ref{sec:test_sequence_summary} of the supplemental material.
For the choices of QPs and quantizer steps $q$, we adopted those specified in the H.264/AVC standard.  

To experimentally measure the PSNR gain, we first connect a set of experimentally obtained R--D points under different QPs using a piecewise linear line for estimating R--D curves before and after reshaping. 
We then use the bitrate that corresponds to the middle of the tested QP range to find the PSNR gain and consider it as the real/measured PSNR gain.
To obtain the theoretically predicted gain, we need the slope ratio $\eta$ and the reshaping range expansion factor $k$ in \eqref{eq:psnr_gain_theory}. 
To estimate $\eta$, we use the estimates for slopes $m_0$ and $m_{0,1}$. 
The expansion factor $k$ is estimated from the supports of the histograms before and after the range expansion.

\subsection{Rate--Distortion Performance and Predicted Gain}

\begin{figure*}[!t]
  \centering
  \subfloat[][]{\includegraphics[clip, trim=1 0 8mm 4mm, width=0.315\linewidth]{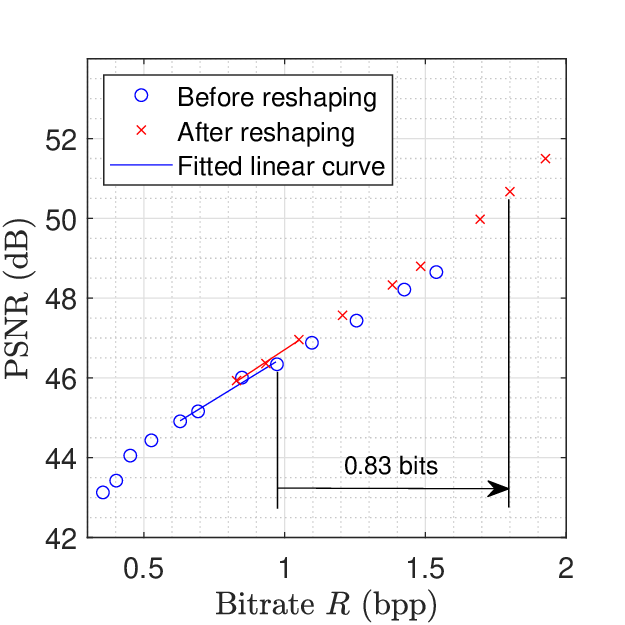}}
  \hspace{2mm}
  \subfloat[][]{\includegraphics[clip, trim=1 2mm 8mm 0, width=0.33\linewidth]{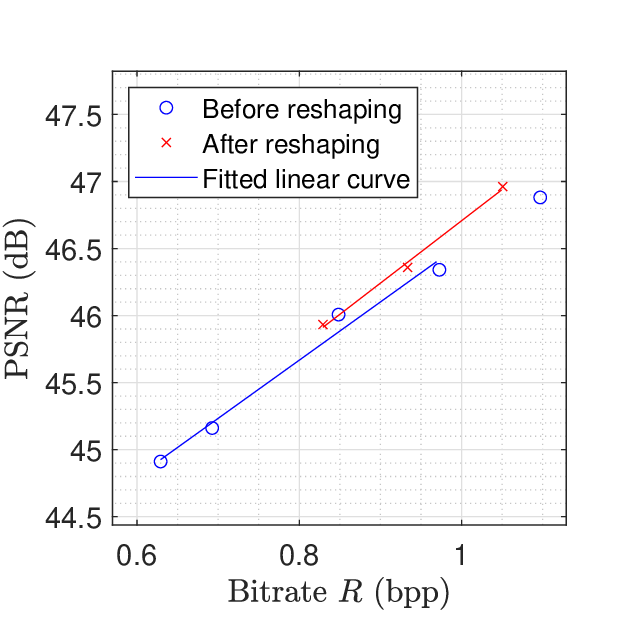}}
  \hspace{2mm}
  \subfloat[][]{\includegraphics[clip, trim=1 0 0mm 0, width=0.315\linewidth]{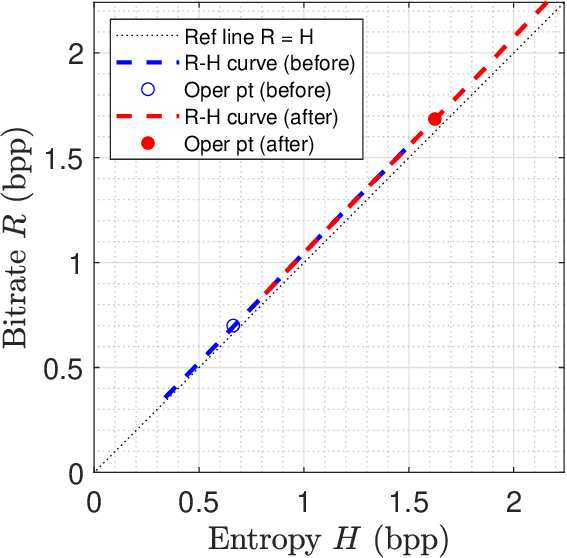}}
  \caption{Example plots from sequence \textsc{BasketballDrill} for measuring (a)--(b)~the empirical PSNR gain and (c)~the quantities needed for predicting the theoretical PSNR gain. 
  (a)~Rate--distortion~(R--D) curves before and after reshaping. Each circle corresponds to a specific quantization step. (b)~Zoomed-in version of~(a). The plots reveal that reshaping causes the R--D curve to move to the top-right corner of the R--D plane, leading to an increased measured/empirical PSNR as predicted by our theoretical result~\eqref{eq:psnr_gain_theory}.
  (c)~Corresponding bitrate--entropy~(R--H) curve, which is contributed by the R--H points before~(blue dashed) and after~(red dashed) reshaping. 
  The R--H curve sits above the diagonal line, confirming the suboptimality of the entropy coder stated in~\eqref{eq:suboptimal_mapping}. The two operating points $(\hat{H}^{(0)}, \hat{R}^{(0)})$~(``$\circ$'') and $(\hat{H}^{(1)}, \hat{R}^{(1)})$~(``$\bullet$'') for predicting the theoretical PSNR gain are annotated on the R--H curve.}
  \label{tab:exp_rd}
\end{figure*}

We use the \textsc{BasketballDrill} sequence as an example in Fig.~\ref{tab:exp_rd} to describe the experimental results and explain the connection to the theoretical result.
Fig.~\ref{tab:exp_rd}(a) reveals that reshaping leads to increases in both PSNR and bitrate, and more importantly, the R--D points after range expansion (in red ``$\times$'') sit to the top-right of the original R--D points (in blue ``$\circ$'').
We annotated the shift of one such operating point.
Specifically, the right arrow in Fig.~\ref{tab:exp_rd}(a) shows that the increase of bitrate due to reshaping is $0.83$~bits, which is less than the largest possible improvement of $\log_2 1.95 = 0.96$~bits in terms of entropy, in which we plug the measured range expansion factor $\hat{k} = 1.95$ into the theoretical results of \eqref{eq:H1_additive}.
Fig.~\ref{tab:exp_rd}(b) zooms into the middle range of Fig.~\ref{tab:exp_rd}(a), which clearly shows that the new R--D curve has a higher PSNR than the original R--D curve at the same bitrate.
The experimental results support a positive coding gain as predicted by our theoretical model.
We note that Fig.~\ref{tab:exp_rd}(a)--(b) experimentally verify the theoretical results in \eqref{eq:subopt-theory-rd}--\eqref{eq:DR-theory} and the simulated results in Fig.~3(b).

Fig.~\ref{tab:exp_rd}(c) shows an R--H curve resulting from encoding the \textsc{BasketballDrill} sequence and its shape validates our assumption in \eqref{eq:suboptimal_mapping} about the suboptimality of entropy coders.
The empirical R--H curve sits above the reference line of $R = H$, which experimentally verifies the theoretical illustration in Fig.~\ref{fig:br_vs_H}.
Such suboptimality revealed by the R--H curve will lead to the gain in coding efficiency as our theory in~\eqref{eq:psnr_gain_theory} predicts.

\begin{figure*}[!t]
  \includegraphics[width=0.32\linewidth]{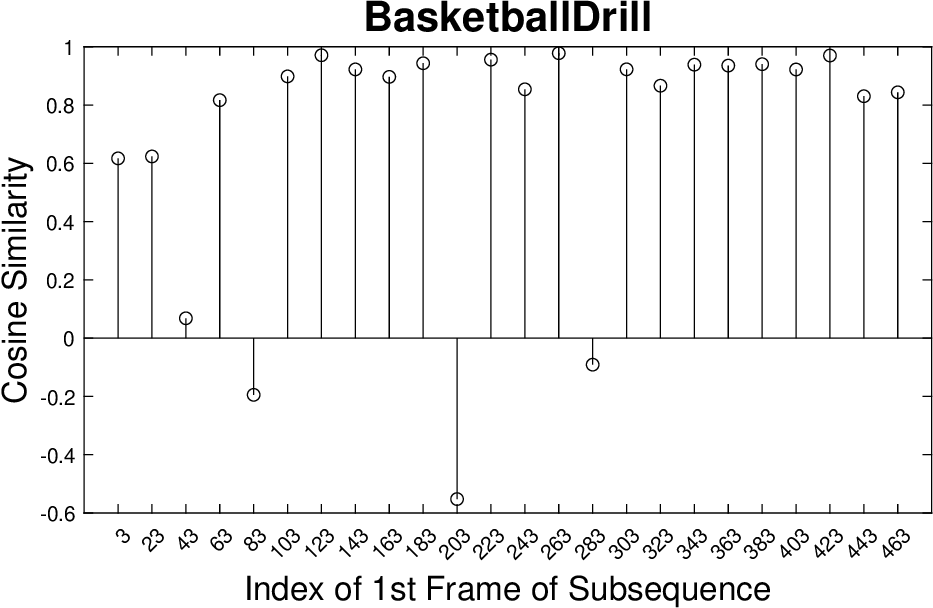}
  \hspace{2mm}
  \includegraphics[width=0.32\linewidth]{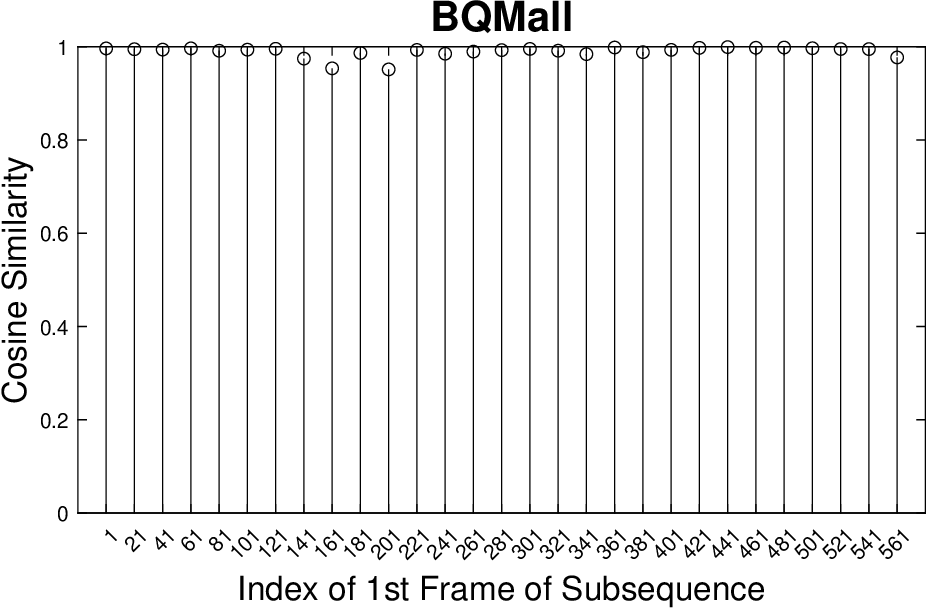}
  \hspace{2mm}
  \includegraphics[width=0.32\linewidth]{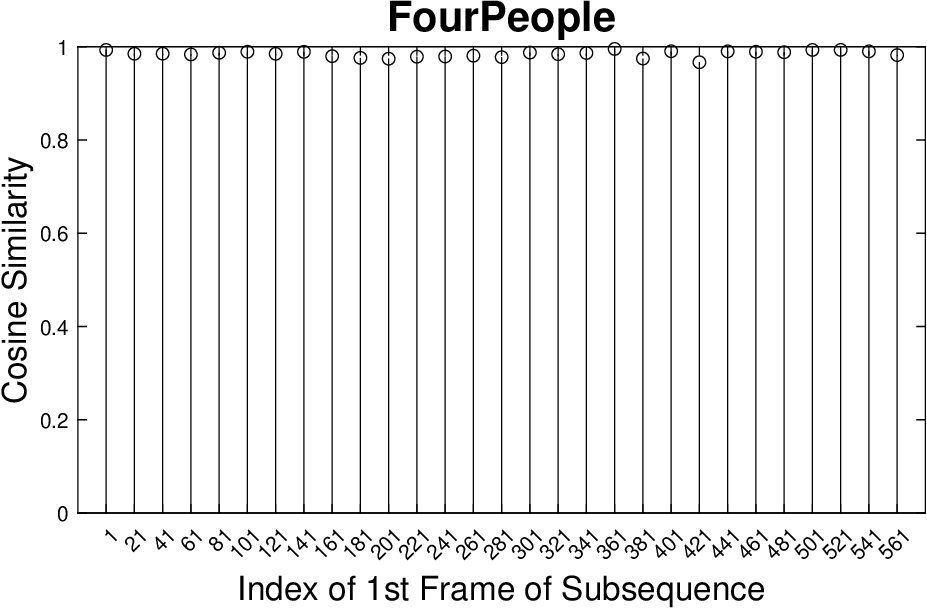} \\
  \subfloat[][]{
    \includegraphics[clip, trim=0 0 0 4mm, height=26.6mm]{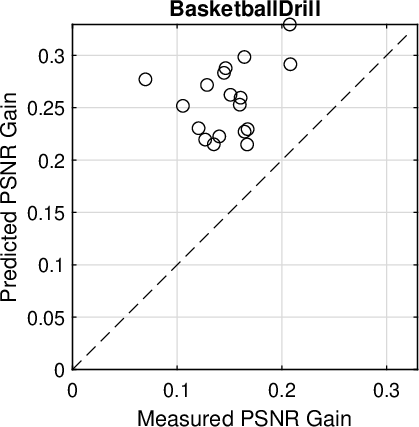}
    \includegraphics[clip, trim=0 0 0 4mm, height=26.6mm]{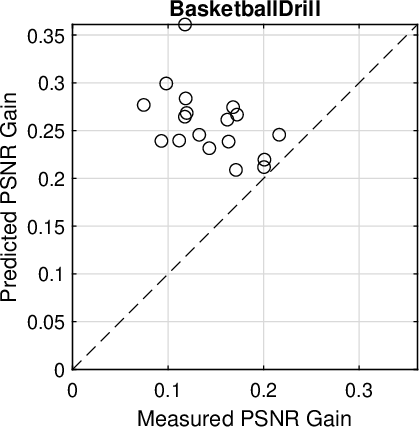}
  }\hspace{1mm}
  \subfloat[][]{
    \includegraphics[clip, trim=0 0 0 4mm, height=26.6mm]{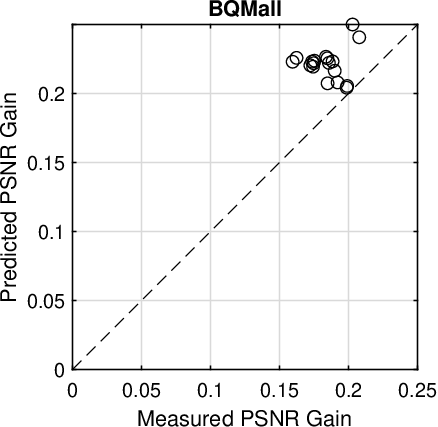}
    \includegraphics[clip, trim=0 0 0 4mm, height=26.6mm]{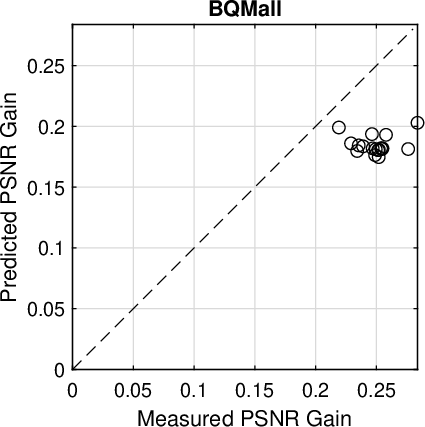}
  }\hspace{1mm}
  \subfloat[][]{
    \includegraphics[clip, trim=0 0 0 4mm, height=26.6mm]{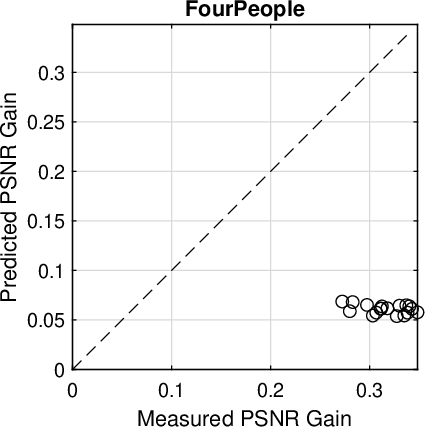}
    \includegraphics[clip, trim=0 0 0 4mm, height=26.6mm]{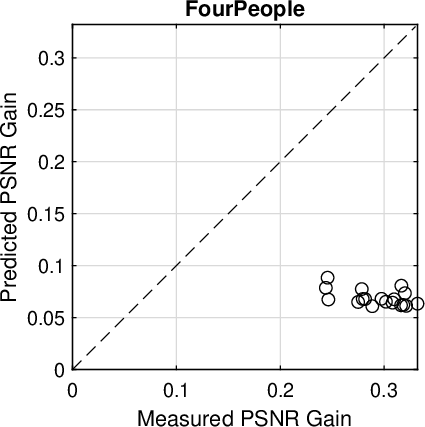}
  } \vspace{4mm}\\ 
  \includegraphics[width=0.32\linewidth]{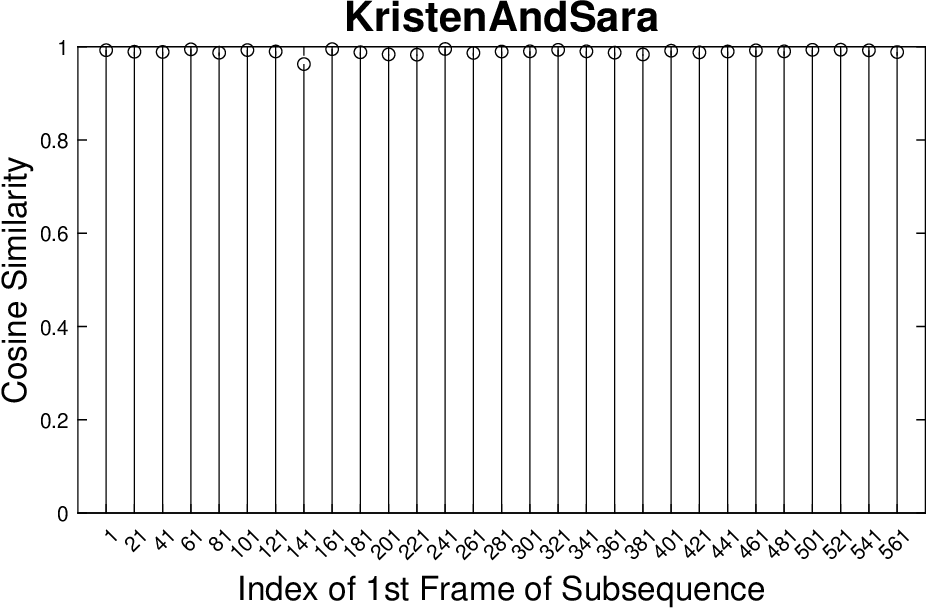}
  \hspace{2mm}
  \includegraphics[width=0.32\linewidth]{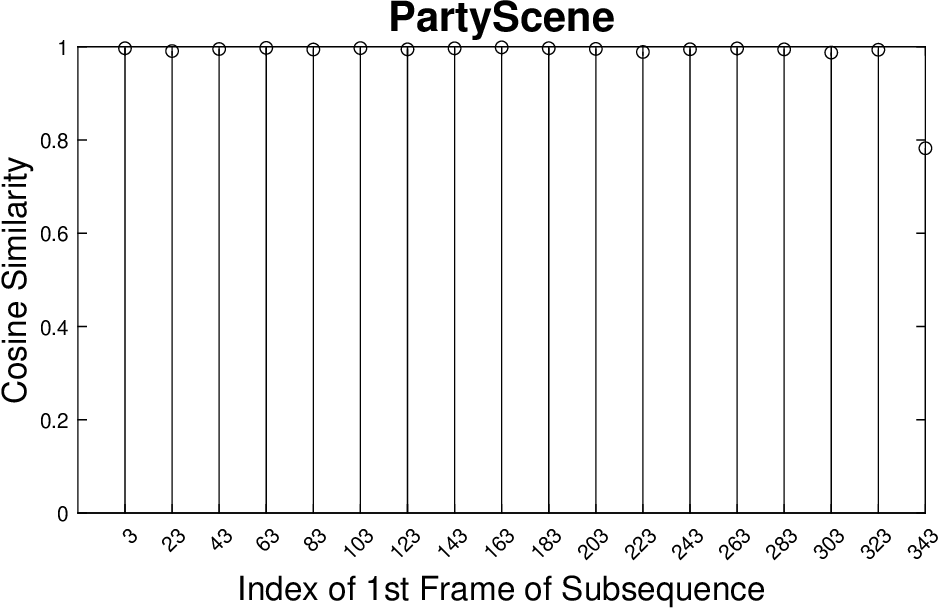}
  \hspace{2mm}
  \includegraphics[width=0.32\linewidth]{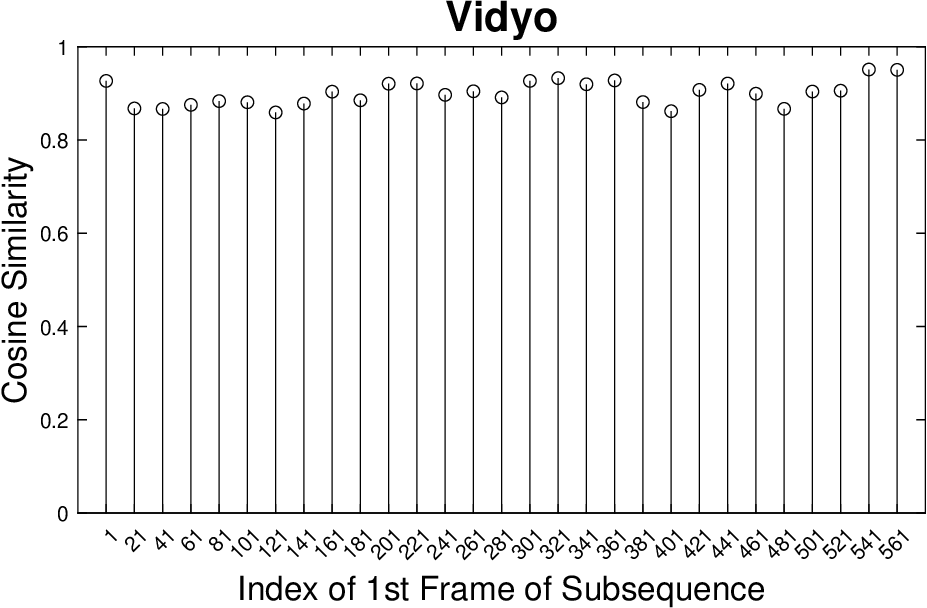}\\
  \subfloat[][]{
    \includegraphics[clip, trim=0 0 0 4mm, height=26.6mm]{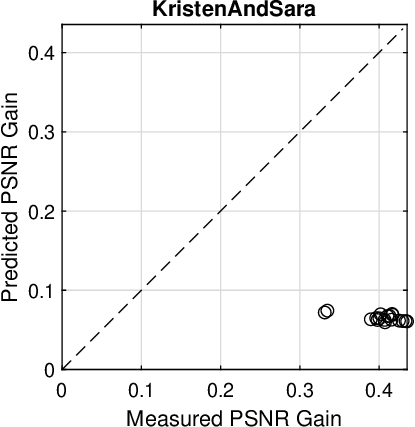}
    \includegraphics[clip, trim=0 0 0 4mm, height=26.6mm]{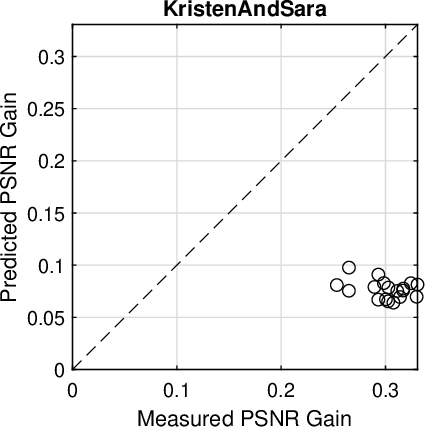}
  }\hspace{1mm}
  \subfloat[][]{
    \includegraphics[clip, trim=0 0 0 4mm, height=26.6mm]{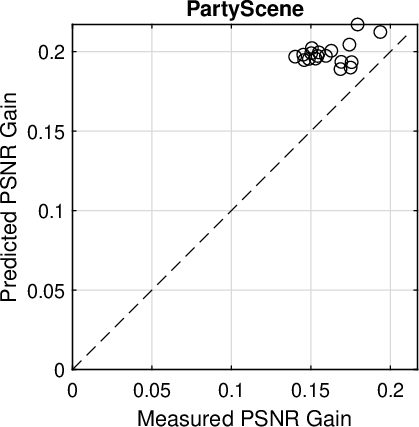}
    \includegraphics[clip, trim=0 0 0 4mm, height=26.6mm]{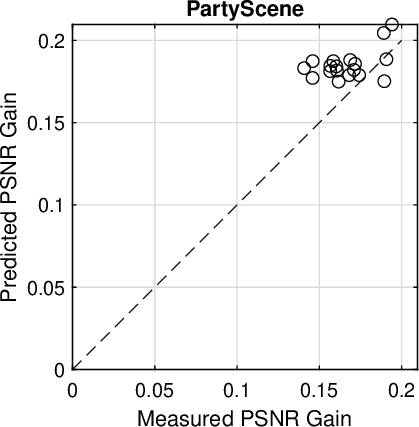}
  }\hspace{1mm}
  \subfloat[][]{
    \hspace{2mm}\includegraphics[clip, trim=0 0 0 4mm, height=26.6mm]{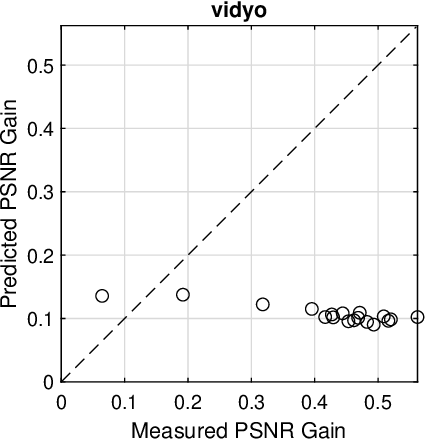}
    \hspace{1mm}\includegraphics[clip, trim=0 0 0 4mm, height=26.6mm]{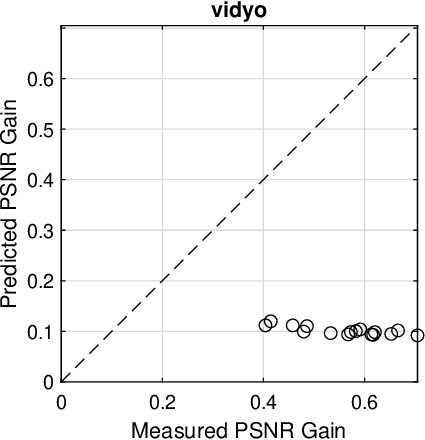}
}
  \caption{Each subfigure from (a) to (f) contains a top plot of cosine similarity values for all test videos' subsequences and two bottom plots of typical scatter plots between measured gain and theoretically predicted gain. Each circle corresponds to the PSNR gain of a frame at a certain bitrate. PSNR gains from the theory and actual measurement are consistently positive in terms of cosine similarity.
  }
\label{fig:scatter}
\end{figure*}

To calculate the slope ratio $\eta$ from the empirically obtained R--H curve in Fig.~\ref{tab:exp_rd}(c), we pick two operating points on the curve. 
First, we use the measured bitrate before reshaping, $\hat{R}^{(0)}$, to find the estimated entropy before reshaping, $\hat{H}^{(0)}$. See the annotated blue circle positioned at $(\hat{H}^{(0)}, \hat{R}^{(0)})$.
Second, the estimated entropy after reshaping calculated as $\hat{H}^{(1)} = \hat{H}^{(0)} + \log_2 \hat{k}$, where $\hat{k}$ is the estimated reshaping range expansion factor. Third, the estimated bitrate after reshaping $\hat{R}^{(1)}$ can then be read from the curve.
See the annotated red dot positioned at $(\hat{H}^{(1)}, \hat{R}^{(1)})$ in Fig.~\ref{tab:exp_rd}(c). 
Finally, an estimate for $\eta$ may be calculated per \eqref{eq:def-eta}, \eqref{eq:m01-def}, and \eqref{eq:slope-m0} as follows:
\begin{equation}
\hat{\eta} = \frac{\hat{R}^{(1)} - \hat{R}^{(0)}}{\hat{H}^{(1)} - \hat{H}^{(0)}} \bigg/ \frac{\hat{R}^{(0)}}{\hat{H}^{(0)}} 
= \left( \frac{\hat{R}^{(1)}}{\hat{R}^{(0)}} - 1 \right) \frac{\hat{H}^{(0)}}{\log_2 \hat{k}}.
\label{eq:eta_estimator}
\end{equation}%
Plugging $\hat{\eta}$ into \eqref{eq:psnr_gain_theory}, we obtain the predicted PSNR gain for one frame.
The process is repeated for a total of 19~P~frames of each subsequence, which results in 19 pairs of theoretically predicted gain and measured gain.
Each video sequence contains 300--400 frames or 15--20 subsequences.
In Fig.~\ref{fig:scatter}, for each of the six testing video sequences, we show two typical scatter plots between predicted gain and measured gain for one subsequence each. More scatter plots can be generated by running our plotting source code.\footnote{Source code for the main program and plotting will be released on the authors' webpages.}
Each circle in a scatter plot corresponds to the PSNR gain of a frame 
in the middle of the tested QP range.
Since all circles fall within the first quad of the 2-D plane, we conclude that for all frames,  PSNR gains from the theory and actual measurement are consistently positive.
To provide a comprehensive evaluation of all subsequences, TABLE~\ref{tab:psnr_gain} further provides quantitative results on the measured PSNR gain and predicted PSNR gain. 
Each cell of measured/predicted gain contains the average and standard deviation of the increased PSNR values from all subsequences of a test sequence.
Even though the predicted PSNR gain may overestimate or underestimate the measured gain, our theoretical model is capable of correctly predicting the sign of the gain for almost all subsequences that were tested.

We also calculate the cosine similarity~\cite{tan2006data} for each subsequence and plot them against the starting index of the corresponding subsequence in Fig.~\ref{fig:scatter}. 
The average and standard deviation of the cosine similarity values of each sequence are given in the right-most column of TABLE~\ref{tab:psnr_gain}.
We note that for 5 of 6 testing sequences, the average cosine similarities are greater than or equal to 0.9.
Even though a small proportion of subsequences in \textsc{BasketballDrill} have smaller cosine similarity,%
\footnote{\textsc{BasketballDrill} used a narrower search range than necessary as a compromise to the computational complexity (see Appendix~\ref{app:search_range} and TABLE~\ref{tab:search_range} of the supplemental material), which made the residues of macroblocks of fast motion abnormally large. This may have caused the abnormality in either/both predicted and measured PSNR gains and therefore lowered the cosine similarity.}
the majority of subsequences in this testing sequence perform as expected.
We therefore confirm that our theoretical model about the PSNR gain is consistent with the experimental results. 
 
\begin{table}[!t]
\renewcommand{\arraystretch}{1.3}
\caption{Measured and Theoretically Predicted PSNR Gains.}
\vspace{-2mm}
\label{tab:psnr_gain}
\centering
\scalebox{0.90}{
\begin{tabular}{lcc|llc}
\toprule
\multirow{2}{*}{\textbf{Sequence}} &  \textbf{Slope} & \textbf{Rate} & \multicolumn{2}{c}{\textbf{PSNR Gain in dB}} & \textbf{Cosine}
\\ 
  & $k$ & (bit) & \textbf{Measured} & \textbf{Predicted} & \textbf{Similarity}
\\ \hline
BasketballDrill & 1.95 & 0.7
& $0.10\,(0.06)^\dag$ & $0.26\,(0.03)$ & $0.70\,(0.43)$ 
\\ \hline
BQMall & 1.95 & 1.4 
& $0.25\,(0.06)$ & $0.18\,(0.03)$ & $0.99\,(0.01)$ 
\\ \hline
FourPeople & 1.93 & 0.8
& $0.30\,(0.02)$ & $0.07\,(0.01)$ & $0.99\,(0.01)$
\\ \hline
Kristen\&Sara & 1.93 & 0.9
& $0.40\,(0.04)$ & $0.06\,(0.01)$ & $0.99\,(0.01)$
\\ \hline
PartyScene & 1.95 & 1.9
& $0.15\,(0.02)$ & $0.19\,(0.01)$ & $0.98\,(0.05)$
\\ \hline
Vidyo & 1.98 & 0.7 
& $0.38\,(0.06)$ & $0.12\,(0.01)$ & $0.90\,(0.03)$
\\ \bottomrule
\multicolumn{6}{l}{$^\dag$ Each reported PSNR gain is averaged over all subsequences and the sample }\\
\multicolumn{6}{l}{ standard derivation is reported within the following parentheses.}
\end{tabular}}
\end{table}

\subsection{Impact of Suboptimal Entropy Coder on Coding Gain}
In this subsection, we experimentally verify that a less optimal entropy coder can lead to a larger coding gain due to reshaping.
We control the degree of suboptimality of the entropy coder by changing the granularity of the codewords of arithmetic coding.
Specifically, in the baseline suboptimal entropy coder that is also used in other sections of this paper, the codewords were quantized with a step size of 100, whereas in ``More Optimal'' and ``Less Optimal'' setups in TABLE~\ref{tab:suboptimal_entropy_coder_impact}, they have quantizing step sizes of 10 and 1000, respectively. 
TABLE~\ref{tab:suboptimal_entropy_coder_impact} shows the measured PSNR gains for each sequence at different suboptimality levels.
The table reveals an increasing trend in coding gain as the entropy coder becomes less optimal, which confirms that reshaping can achieve more gain for entropy coders that are less optimal.

\begin{table}[!t]
\renewcommand{\arraystretch}{1.3}
\caption{Measured Coding Gain for Suboptimal Entropy Coder.}
\vspace{-2mm}
\label{tab:suboptimal_entropy_coder_impact}
\centering
\scalebox{0.90}{
\begin{tabular}{lcccc}
\toprule
\multirow{2}{*}{\textbf{Sequence}} &  \multicolumn{3}{c}{\textbf{Measured PSNR Gain in dB}}\\ 
& More Optimal & Suboptimal & Less Optimal\\ \hline
BasketballDrill & 0.097 & 0.097 & 0.318$^*$\\ \hline
BQMall & 0.246 & 0.248 & 0.280\\ \hline
FourPeople & 0.297 & 0.299 & 0.345\\ \hline
Kristen\&Sara & 0.391 & 0.395 & 0.455\\ \hline
PartyScene & 0.172 & 0.172 & 0.175\\ \hline
Vidyo & 0.379 & 0.384 & 0.612\\ \bottomrule
\multicolumn{4}{l}{$^*$ This entry is obtained when quantizing step size is set to 500 rather than }\\
\multicolumn{4}{l}{the default value of 1000 due to stability issues in implementation.}
\end{tabular}}
\end{table}

\section{Discussion} \label{sec:discuss}
\subsection{Extension to Reshaping Functions with Multiple Pieces}
This paper has investigated the class of reshaping functions that have only one piece of a linear segment with a nonzero slope as shown in Fig.~\ref{fig:reshape_map}. 
In practice, reshaping functions with multiple pieces of linear segments are adopted in the VVC/H.266 standard. 
Our theoretical analysis of MSE and entropy is amendable to the multiple-segment scenario.
See Appendix~\ref{app:twopiece} for pathways for the extension.
However, as the number of segments increases, boundary analysis starts to become complicated, which lowers the mathematical tractability and may make the theoretical result less elegant. 
Nevertheless, it is possible to use simulated data to validate the phenomenon that we have derived and observed in the one-piece scenario. 

\subsection{Mild Quantization Effect in the Codomain of Reshapers}
In the actual implementation of the reshaper in the VVC/H.266 standard, reshapers $g$ and $g^{-1}$ are implemented by lookup tables that generate integer-valued output.
These integer-valued outputs exhibit a mild quantization effect. 
To simplify our investigation, this paper skips such an effect.
However, we note that quantization should have a mild impact on the theoretical results. 

\subsection{Choice of Evaluation Metrics}
\label{ssec:dis_metric}
We choose the cosine similarity~\cite{tan2006data}
as the evaluation metric to verify the proposed theoretical model via the experiments. 
The cosine similarity focuses on the consistency between the theoretically predicted gain and the experimentally measured gain. 
This is in line with this paper's goal of providing a theoretical justification for the performance enhancement achieved through the adoption of the in-loop reshaper.
We did not use metrics that are sensitive to the specific magnitudes of gains or losses, including MSE, mean absolute error~(MAE), and Pearson's correlation~\cite{tan2006data}; instead, we prioritize a metric that can consistently distinguish between gains and losses.
Although being able to predict the precise magnitudes of gains is advantageous, it can be technically challenging to accomplish, considering the complex architecture of video codecs. 
Another drawback of using  Pearson's correlation as the evaluation metric is that its score discards the average signal level when calculating the similarity of two signals being compared~\cite[Ch.~2.4.5]{tan2006data}. In contrast, cosine similarity measures the similarity by considering the context given by the average signal level, which is the achieved PSNR gains' average that our study focuses on.

We also note that other metrics such as Spearman's correlation and Kendall's correlation~\cite[Ch.~2]{kendall1990rank} are unsuitable for the objectives of our work.
Spearman measures the strength and direction of association between two ranked variables, but the exact rank of a predicted gain is not the key to verifying our theoretical result.
Kendall's correlation is also a rank-based metric so it is less relevant.

\subsection{Theoretical Result's Implications and Connections to Real-World Video Codecs}
\label{ssec:limitations}

This theoretical investigation focuses on a general hybrid video coding setup and is not tied to any specific video coding standard such as VVC/H.266.
A key finding was that in-loop reshaping does not yield any rate--distortion improvements when using a hypothetical optimal entropy coder~[see Fig.~\ref{fig:mse_vs_H}(a) and Section~\ref{subsec:no_gain}]. However, for suboptimal entropy coders typically used in practice, we were able to derive the rate--distortion gain in a closed form~[See Eq.~\eqref{eq:psnr_gain_theory}] and confirmed this through experimental validation. We chose to implement one of the simplest codecs to isolate the effects of reshaping from other potential contributing factors.

Reshaping's contribution to enhancing the coding efficiency is largely independent of various predictive coding and transform coding tools found in modern video codecs, such as block partitioning, pyramid inter-frame prediction structures, and mode-adaptive intra-frame prediction. 
The contribution of reshaping $(q/k)^2$ to MSE in \eqref{eq:raw_mse_final} is uncoupled from the contributions in $\left\| e ( \T\r_t / q ) \right\|^{2}$ resulting from the potential use of sophisticated predictive and transform coding tools.
Similarly, reshaping's impact on bitrate, as illustrated in \eqref{eq:r1_decomposition}, is additive to the residue's entropy $H^{(0)}$.
From the perspective of information theory, all these modern coding tools aim to reduce redundancies in residues.
However, our reshaping analysis does not primarily concern the residues; instead, it concerns the suboptimality (see Fig.~\ref{fig:mse_vs_H}) of the entropy coder that compresses the residues.

While our theoretical result applies generally to hybrid video coding, the exact amount of coding gain derived from our mathematical model cannot be directly used to predict a real-world codec's gain. If a reader wishes to examine the reshaping coding gain for an actual codec, one possibility is to restrict the codec's configuration to that of our simplified codec, i.e., using fixed block and transform sizes. Otherwise, the reader must expand the theoretical analysis to align with the codec's specific configuration. The current distortion analysis covers general cases for adaptive transform sizes but does not address scenarios involving multiple QPs within a frame. The entropy analysis also requires expansion to accommodate various block-partitioning and transform sizes, as different sizes result in different residual characteristics whose bitrates need to be separately calculated.

\section{Conclusion} \label{sec:conclude}
In this paper, we have theoretically analyzed the rate--distortion performance of the in-loop reshaper and have proved that coding efficiency can be improved only when the entropy coder is suboptimal---the scenario that video coding practically deals with.
We have derived the PSNR gain in the closed form and have shown that the theoretically predicted gain is consistent with that measured from experiments using 
a simplified codec and standard testing video sequences.
We note that if real data and codec depart from the modeling assumptions mentioned in this work, coding gain due to reshaping could be affected by terms such as $q$, $n$, $m_0$, and $h^{(0)}$ involved in the rate--distortion analysis and/or unaccounted factors such as motion compensation.

\appendices
\section{Justification to Eq.~\eqref{eq:full_pipeline_eq_final}}
\label{app:derivation_recon_err}
This section theoretically justifies the approximation sign that connects~\eqref{eq:pipeline_eq6} and~\eqref{eq:full_pipeline_eq_final}. For readers' convenience, we restate the problem as follows:
\begin{subequations}
\begin{align}
\hatIt 
&= g^{-1} \Big( g(\I_t) + \T^{-1} q \cdot e\big( \T\r_t / q \big) \Big) \label{eq-app:hat_I_t_step1}\\
&\approx \I_{t} - k^{-1} \T^{-1} q \cdot e \big( \T\r_t / q \big).
\label{eq-app:hat_I_t_step2}
\end{align}%
\end{subequations}%
It is equivalent to show that the unconditional expected reconstruction error
\begin{equation}
\E \Big[ \big(\hat{I}_t(i) - I_t(i) \big)^2 \Big] \approx \left(\frac{q \sigma_u}{k}\right)^2 = \frac{(q/k)^2}{12},
\label{eq-app:proof-obj}
\end{equation}
where $i \in \{1, \dots, N\}$ is the coordinate of each vector.

We will show that unconditional expected reconstruction error is composed of three components: (\textit{i})~the errors due to clipping of reconstructed pixel values $\hat{I}_t(i)$ from below at $0$, (\textit{ii})~the errors due to quantization when $\hat{I}_t(i) \in [0, M]$, and (\textit{iii})~the errors due to clipping of  $\hat{I}_t(i)$ from above at $M$.
We will see that the second component dominates the unconditional expected reconstruction error and the other two components are negligible.

We start the analysis from the vector form and will eventually force on the $i$th coordinate.
We first examine the column vector $g(\I_t) + \boldsymbol{\epsilon}$ inside the backward reshaping function $g^{-1}$ in \eqref{eq-app:hat_I_t_step1}, where we define $\boldsymbol{\epsilon} = \T^{-1} q \cdot e\big( \T\r_t / q \big)$. 
We note that $\e = e\big( \T\r_t / q \big)$ is a random vector with each coordinate $e_i$ uniformly distributed in $[-1/2, 1/2]$. The forward transform $\T$ decorrelates $\r_t$, making $\{e_i\}$ nearly uncorrelated.
We denote the $i$th element of vector $g(\I_t) + \boldsymbol{\epsilon}$ to be $g(I_t(i)) + \epsilon_i$, where $\epsilon_i = q \sum_{j=1}^N t_{ij} e_j$, where $t_{ij}$ is the $(i,j)$th element of $\T^{-1}$. 
We calculate $\E[\epsilon_i] = 0$ and $\var(\epsilon_i) = q^2 \sum_{j=1}^N t_{ij}^2 \var(e_j) = q^2 \sigma_u^2$, where $\sigma_u^2 = \var(e_j) = 1/12$. We use the property that $\{e_i\}$ are uncorrelated and $\sum_{j=1}^N t_{ij}^2 = 1$ for the orthogonal transform $\T^{-1}$.
Using the central limit theorem, we approximate $\epsilon_i \sim \mathcal{N}(0, q^2 \sigma_u^2),\ \forall i$.

We now analyze the errors due to clipping of reconstructed pixel values $\hat{I}_t(i)$ from below at $0$. Formally, we define an event $E_1 = \{g(I_t(i)) + \epsilon_i < 0\}$. When $E_1$ is true, the backward reshaper has to return $g^{-1}(\cdot) = a$. 
We now evaluate how often this will happen:
\begin{subequations}
\begin{align}
\P[E_1] &= \P \left[ \phi_i + \epsilon_i < 0 \right] \\
&= \E \big[ \P [\epsilon_i < -\phi_i \mid \phi_i ] \big] \\
&\approx  \E \left[ \Phi\left( \tfrac{-\phi_i}{q \sigma_u} \right) \right] \\
&= \sum_y \Phi\left( \tfrac{-\phi_i}{q \sigma_u} \right) p_{\phi_i}(y) \\
&= \sum_{x=a}^b \Phi\left( \tfrac{-k(x - aM)}{q \sigma_u} \right) p_I(x), \label{eq-app:pr-E1}
\end{align}
\end{subequations}
where $\phi_i = g(I_t(i))$ is defined to simplify the notation, $\Phi(\cdot)$ is the CDF of standard Gaussian, and $p_X(x)$ is the probability mass function (PMF) of any random variable $X$.
We then evaluate the expected reconstruction error when clipping from below happens: 
\begin{subequations}
\begin{align}
&\quad\ \E\left[ \big( \hat{I}_t(i) - I_t(i) \big)^2 \Big| E_1 \right] \notag \\
&= \sum_{x=a}^b \E \left[ \text{err}(i)^2 \mid I_t(i) = x, E_1 \right] \P[I_t(i) = x \mid E_1]\\
&= \sum_{x=a}^b (x - a)^2 \, \P[E_1 | I_t(i) = x] \, p_I(x) \, \big/ \, \P[E_1]\\
&= \sum_{x=a}^b (x - a)^2 \, \Phi\left( \tfrac{-k(x-aM)}{q \sigma_u} \right) p_I(x) \, \big/ \, \P[E_1], \label{eq-app:expected-err-E1}
\end{align}
\end{subequations}
where $\text{err}(i) = \hat{I}_t(i) - I_t(i)$.
Multiplying \eqref{eq-app:pr-E1} and \eqref{eq-app:expected-err-E1}, we obtain the error contribution from $E_1$.

We can similarly analyze the errors due to clipping of reconstructed pixel values $\hat{I}_t(i)$ from above at $M$. Formally, we define another event $E_3 = \{g(I_t(i)) + \epsilon_i > M\}$.
When $E_3$ is true, the backward reshaper has to return $g^{-1}(\cdot) = b$. 
Let us assume that the input distribution is symmetric, i.e., $a = M - b$. This allows us to directly reuse the result from $E_1$.

We now analyze the last of the three cases, i.e., errors due to quantization when $\hat{I}_t(i) \in [0, M]$. Formally, we define a third event $E_2 = \{0 < g(I_t(i)) + \epsilon_i < M\}$.
When $E_2$ is true the backward reshaper operates normally in the segment of $g^{-1}(x) = x/k + aM$ that directly inverts to obtain a close approximation of the original input:
\begin{subequations}
\begin{align}
\hat{I}_t(i) &= g^{-1} \big(g(I_t(i)) + \epsilon_i \big) \\
&= \left[ k(I_t(i) - aM) + \epsilon_i \right] / k + aM\\
&=  I_t(i) + k^{-1} \epsilon_i.
\end{align}
\end{subequations}
Hence, 
\begin{equation}
\E \left[ \big(\hat{I}_t(i) - I_t(i)\big)^2 \Big| E_2 \right] = \E \left[ (k^{-1} \epsilon_i)^2 \right] = (q/k)^2 \sigma_u^2.    
\end{equation}
Since $E_1$, $E_2$, and $E_3$ are mutually exclusive and union to the sample space, we have $\P[E_2] = 1 - 2 \P[E_1]$.

Putting all three cases together, the unconditional expected reconstruction error is
\begin{subequations}
\begin{align}
&\ \E \left[ (\hat{I}_t(i) - I_t(i))^2 \right] \notag \\
=&\, \sum_{k=1}^3 \E\left[ \big( \hat{I}_t(i) - I_t(i) \big)^2 \Big| E_k \right] \P[E_k] \\
\approx
&\, \left(\frac{q \sigma_u}{k}\right)^2 
\left[ 
1 - 2 \sum_{x=a}^b \Phi\left( \tfrac{-(x - aM)}{q \sigma_u / k} \right) p_I(x)
\right] \notag \\
&\, +2 
\sum_{x=a}^b (x - a)^2 \, \Phi\left( \tfrac{-(x-aM)}{q \sigma_u / k} \right) p_I(x),
\end{align}
\end{subequations}
where the equation contains a finite number of error functions that are close to $0$.
We conducted a simulation using $M = 1,\!023$, $a \in \{100, \dots, 280\}$, a uniform $p_I(x)$, and $q \in \{5, 10, 20, 40, 80, 160, 224\}$. The results in TABLE~\ref{tab:mse_precision} reveal that \eqref{eq-app:proof-obj} holds with high precision. For example, at $\text{QP} = 30$, the largest percentage error is only $0.26\%$.
\begin{table}[!t]
\renewcommand{\arraystretch}{1.3}
\caption{Precision of Approximated Reconstruction Error.\vspace{-2mm}}
\label{tab:mse_precision}
\centering
\begin{tabular}{ccc}
\toprule
QP & QStep & Error Range \\ \hline
18 & 5 & $[0.11\%, 0.18\%]$ \\ \hline
24 & 10 & $[0.14\%, 0.19\%]$ \\ \hline
30 & 20 & $[0.21\%, 0.26\%]$ \\ \hline
36 & 40 & $[0.36\%, 0.41\%]$ \\ \hline
42 & 80 & $[0.67\%, 0.71\%]$ \\ \hline
48 & 160 & $[1.28\%, 1.32\%]$ \\ \hline
51 & 224 & $[1.77\%, 1.82\%]$ \\ \bottomrule
\end{tabular}
\end{table}
\section{Differential Entropy and Proof of Range Expansion Gain}
\label{sec-app:didff-entropy}

\subsection{Differential Entropy Review}
\label{sec:didff-entropy-background}
Let us denote a quantized random variable $X^q$ obtained from quantizing a continuous random variable $X$ with step $q$. 
The differential entropy of the continuous random variable $X$ is defined using its probability density function~(PDF) $f_X(x)$~\cite[Ch.~8]{cover2006elements} as follows:
\begin{equation}
h(X) =
-\int_{-\infty}^{\infty} f_X(x) \log_2 f_X(x) \, dx,
\label{eq:diff_entropy}
\end{equation}
which is a fixed number for a given PDF with constant parameters.
The discrete entropy of the corresponding quantized random variable $X^q$ can be approximated by the differential entropy of $X$ and an offset quantity related to the quantization step~\cite[Thm.~8.3.1]{cover2006elements}, namely,
\begin{equation}
H(X^q) \approx h(X) - \log_2{q},
\label{eq:H_h_relation}
\end{equation}
where $H(X^q)$ is the discrete entropy of $X$ after being quantized with step $q$. 
The theorem reveals that the entropy is determined by (\textit{i}) the shape of the PDF of $X$, i.e., the fixed number $h(X)$ when the PDF of $X$ is given, and (\textit{ii}) and the accuracy of the quantization, i.e., the number of bits needed to represent the uniformly spaced levels in a unit interval.
The approximation in (\ref{eq:H_h_relation}) becomes more accurate as the quantizer becomes finer and it becomes exact as $q \rightarrow 0$~\cite[Thm.~8.3.1]{cover2006elements}.
This theoretical insight is consistent with the observation that more quantized levels lead to larger discrete entropy $H(X^q)$.

\subsection{Proof of Entropy Increase Due to Range Expansion}
\label{sec-app:proof-range-expan-boost-entro}
Let us denote one of the coordinates of transformed residues before and after reshaping by $X$ and $Y$, and their PDFs by $f_X(x)$ and $f_Y(y)$, respectively. Let us assume that raw residue $X \in [-c, c]$. 
Given that $Y = kX$, and we apply the formula for random variable transformation~\cite[Ch.~4.7]{grimmett2020probability} to relate the two PDFs as follows:
\begin{equation}
f_{Y}(y)
=k^{-1} f_{X}\left(y/k\right).
\end{equation}
The differential entropies before and after reshaping are
\begin{subequations}
\begin{align}
h^{(0)} &= -\int_{-c}^c f(x) \log_2 f(x) dx,
\label{eq:h0_final} \\
h^{(1)} &= -\int_{-kc}^{kc} k^{-1} f\left(y/k\right) \log_2 \left[k^{-1} f(y/k)\right] dy\\
&=-\int_{-c}^c f(\xi) \log_2 \left[\frac{1}{k} f(\xi) \right] d \xi, \label{eq:h1_final}
\end{align}
\end{subequations}
where $f(x)$ denotes $f_X(x)$ for simplicity and
(\ref{eq:h1_final}) is obtained from a change of variable, $\xi \leftarrow y/k$.
The change of the differential entropy due to reshaping is therefore: 
\begin{equation}
h^{(1)}-h^{(0)}=-\int_{-c}^{c} f(x) \log_2 \left(\frac{1}{k}\right) d x=\log_2 k>0,
\label{eq:diff_entropy_change}
\end{equation}
where we assume $k>1$, the range expansion scenario.

The theorem in (\ref{eq:H_h_relation}) reveals that the difference between differential entropy converges to the difference between entropy as the quantizer becomes finer, namely, 
\begin{equation}
h^{(1)}-h^{(0)} \approx H^{(1)}-H^{(0)}.
\end{equation}
Together with (\ref{eq:diff_entropy_change}), we obtain 
\begin{equation}
H^{(1)} \approx H^{(0)}+\log_2 k.
\label{eq-app:H1_additive}
\end{equation}
Hence, we conclude that the range expansion due to reshaping increases the entropy by roughly $\log_2 k$.

\section{Pathways for Extending Analysis to Multiple Pieces}
\label{app:twopiece}

\begin{figure}[!t]
  \centering
  \includegraphics[width=0.8\linewidth]{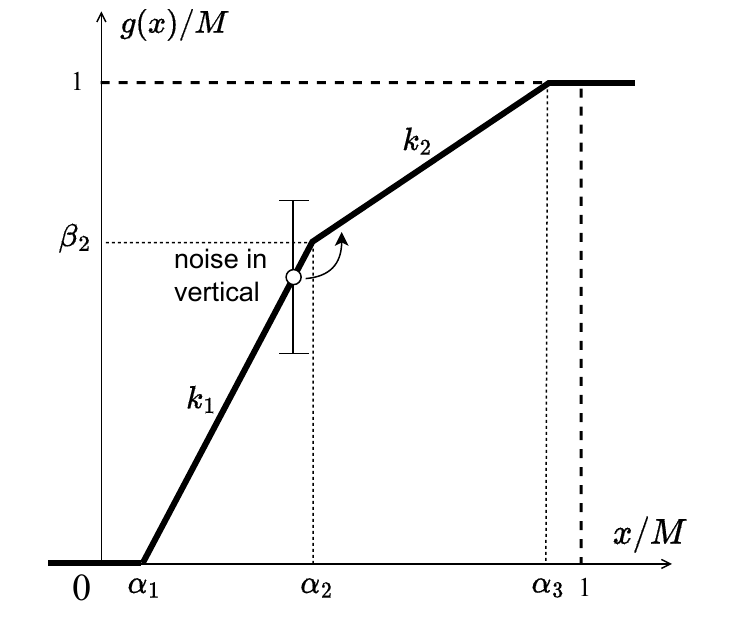}
  \caption{A two-piece forward reshaping function. Segment 1 has a slope of $k_1$ and segment 2 has a slope of $k_2$. A pixel with intensity less than $\alpha_2$ on segment 1 will be corrupted by quantization noise and may be reconstructed to have intensity greater than $\alpha_2$, leading to a larger reconstruction error. This ``crosstalk'' is new to two-piece reshaping functions and is likely more severe in multi-piece functions.}
  \label{fig:twopiece}
\end{figure}

We provide comments and preliminary results for potentially extending the analysis to multi-piece functions.
We use a two-piece example as follows
\begin{align}
g(x) = 
\begin{cases}
0, & x \le \alpha_1 M,\\
k_1(x - \alpha_1 M), & \alpha_1 M < x \le \alpha_2 M, \\
k_2(x - \alpha_3 M) + M, & \alpha_2 M < x \le \alpha_3 M, \\
M, & x > \alpha_3 M.
\end{cases}
\end{align}
and plot it Fig.~\ref{fig:twopiece}.
We approach the reconstruction error and the bitrate separately.

\subsection{Unconditional Expected Reconstruction Error}
The unconditional expected reconstruction error for the two-piece scenario can still be decomposed into the three cases as the one-piece scenario analyzed in Appendix~\ref{app:derivation_recon_err}.
The following two cases can be approached the same way as in one piece, i.e., 
the case $E_1$ (in Appendix~\ref{app:derivation_recon_err}) of having errors due to clipping of reconstructed pixel values $\hat{I}_t(i)$ from below at $0$, and the case $E_3$ of having errors due to clipping of $\hat{I}_t(i)$ from above at $M$.
However, the case $E_2$ of having errors due to quantization when $\hat{I}_t(i) \in [0, M]$ will still need to be split into three subcases.

The first subcase $E_2^{1 \to 2} = \{ g(I_t(i)) < \beta_2 < g(I_t(i)) + \epsilon_i\}$ corresponds to input pixel intensities $I_t(i) \in [\alpha_1, \alpha_2]$ being reconstructed in $(\alpha_2, \alpha_3]$. See the white dot in Fig.~\ref{fig:twopiece} as an example. 
Similarly, the second subcase $E_2^{2 \to 1} = \{ g(I_t(i)) + \epsilon_i < \beta_2 < g(I_t(i)) \}$
corresponds to input pixel intensities $I_t(i) \in (\alpha_2, \alpha_3]$ being reconstructed in $[\alpha_1, \alpha_2]$.
In both subcases of reconstruction onto the neighboring segment, the reconstruction errors may be larger or smaller than the reconstruction on the original piece and are mathematically tractable.
When more pieces are used for the reshaping function or a larger quantization step is used, the chance of having these ``crosstalk'' events is higher, e.g., a larger $\P[E_2^{1 \to 2}]$. It is also possible that more pieces will reduce the reconstruction error, e.g., a smaller $\E[ ( \hat{I}_t(i) - I_t(i) )^2 \mid E_2^{1 \to 2} ]$. We encourage interested readers to look into the combined effect.

The last subcase is the $E_2^{\text{rest}} = \{0 < g(I_t(i)) + \epsilon_i < M\} \text{\textbackslash} (E_2^{1 \to 2} \cup E_2^{2 \to 1})$, whose reconstruction error has the same simple form as $E_2$ of the one-piece scenario.

\subsection{Entropy and Bitrate}
On the entropy front, because the transformed residues are zero mean and close to Gaussian (we provide plots in simulation code) due to the central limit theorem, one can calculate the entropy gain by examining the change in the variance of the transformed residues. Our derivation reveals that the entropy gain after range expansion can be generalized from the one-piece result $\log_2 k$ to the two-piece result
$\log_2 \left( w_1 k_1^2 + w_2 k_2^2 \right)^{1/2}$,
where $w_1 = \int_{\alpha_1 M}^{\alpha_2 M} f_X(x) dx$ is left portion [integrated up to $(100\alpha_2)\% $] of the probability mass and $w_2$ is the right portion. It implies that in addition to the slope parameters $k_1$ and $k_2$ as in \eqref{eq:psnr_gain_theory}, input signal statistics such as $w_1$ and $w_2$ may also affect the rate--distortion performance.

\bibliographystyle{IEEEtran}
\bibliography{inloop_reshaping}
\vskip 0pt plus -1fil
\begin{IEEEbiography}[{\includegraphics[width=1in,height=1.25in,clip,keepaspectratio]{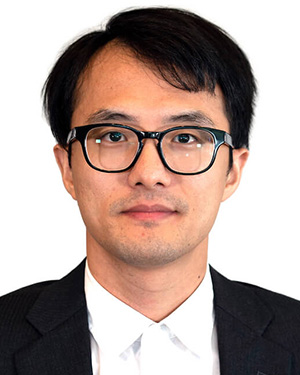}}]{Chau-Wai Wong}
(Member, IEEE) received his B.Eng. and M.Phil. degrees in electronic and information engineering from The Hong Kong Polytechnic University, in 2008 and 2010, and his Ph.D. degree in electrical engineering from the University of Maryland, College Park, MD, USA, in 2017. He is currently an Assistant Professor with the Department of Electrical and Computer Engineering, the Forensic Sciences Cluster, and the Secure Computing Institute, at North Carolina State University. He was a data scientist at Origin Wireless, Inc., Greenbelt, MD, USA. His research interests include multimedia forensics, statistical signal processing, machine learning, data analytics, and video coding. Dr. Wong received an NSF CAREER Award, Top-Four Student Paper Award, HSBC Scholarship, and Hitachi Scholarship. He was involved in organizing the third edition of the IEEE Signal Processing Cup in 2016 on electric network frequency forensics.
\end{IEEEbiography}

\vskip 0pt plus -1fil
\begin{IEEEbiography}[{\includegraphics[width=1in,height=1.25in,clip,keepaspectratio]{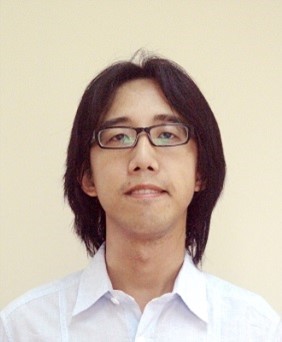}}]{Chang-Hong Fu}
(Member, IEEE) received the B.Eng. (Hons.) and Ph.D. degrees from The Hong Kong Polytechnic University in 2002 and 2008, respectively. He was a Postdoctoral Research Fellow at The Hong Kong Polytechnic University from 2008 to 2011. He joined Nanjing University of Science and Technology in 2011 as an Associate Professor in the School of Electronic and Optical Engineering. He has authored and coauthored over 50 research articles in international journals and conferences. His current research and technical interests include multimedia technologies, video compression, future video coding standards, video-based vital sign monitoring, and multi-sensor fusion for healthcare. Dr. Fu was the Publication Chair of the 2019 IEEE 2nd International Conference on Multimedia Information Processing and Retrieval.
\end{IEEEbiography}

\vskip 0pt plus -1fil
\begin{IEEEbiography}[{\includegraphics[width=1in,height=1.25in,clip,keepaspectratio]{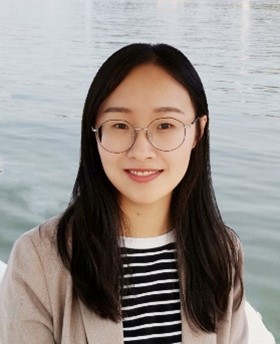}}]{Mengting Xu}
received her B.Eng. degree and M.S. degree from Nanjing University of Science and Technology in 2018 and 2021, respectively. Her research interests include 3D video coding, image signal processing, and deep learning in video coding.
\end{IEEEbiography}

\vskip 0pt plus -1fil
\begin{IEEEbiography}[{\includegraphics[width=1in,height=1.25in,clip,keepaspectratio]{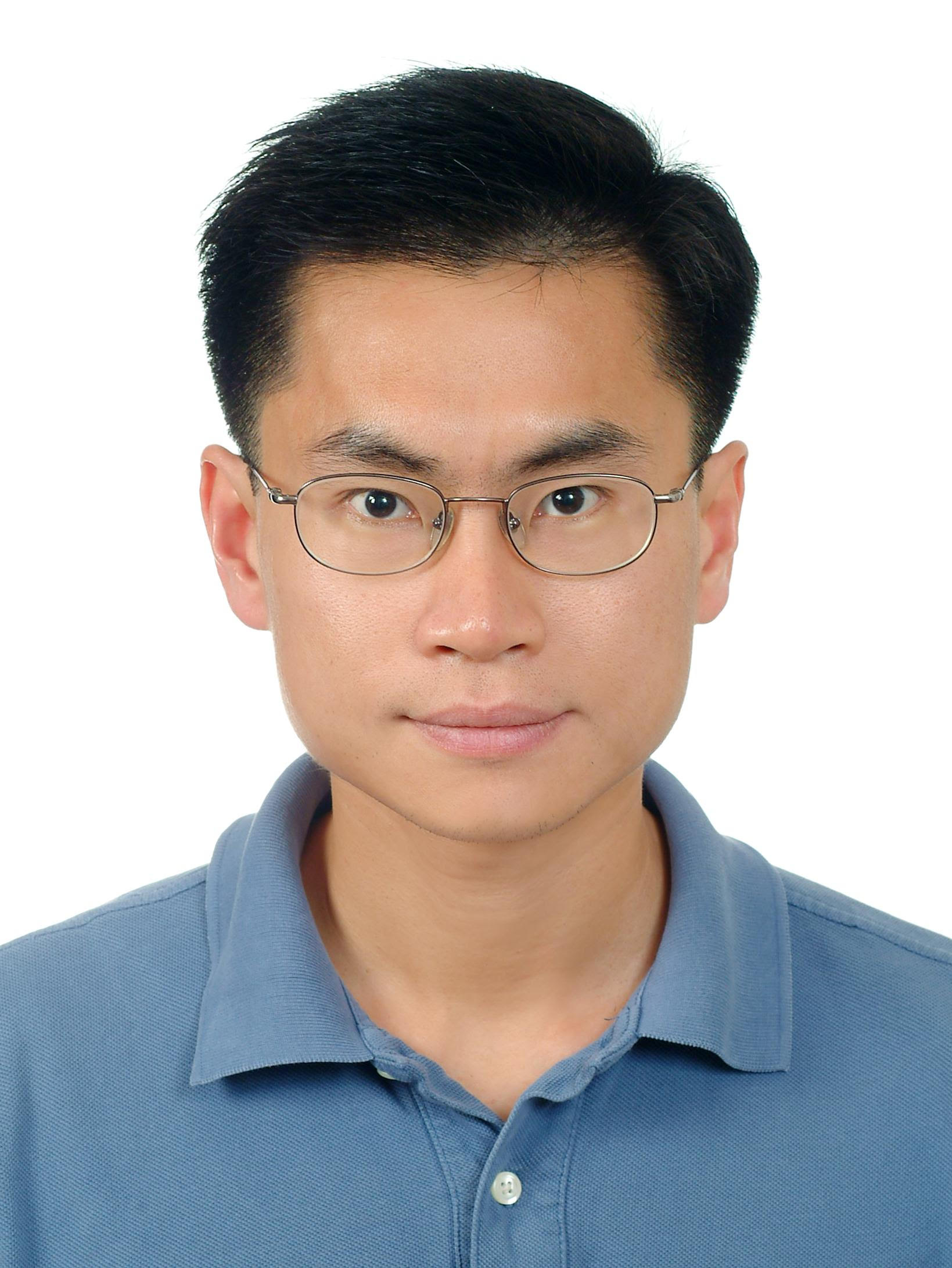}}]{Guan-Ming Su}
(Senior Member, IEEE) obtained his Ph.D. degree from the University of Maryland, College Park.
He is currently the Director of Image Tech Dev at Dolby Labs, Sunnyvale, CA, USA. Prior to this, he has been with the R\&D Department, Qualcomm, Inc., San Diego, CA; ESS Technology, Fremont, CA; and Marvell Semiconductor, Inc., Santa Clara, CA. He is the inventor of 180+ U.S./international patents and pending applications. He served as an associate editor in Asia Pacific Signal and Information Processing Association~(APSIPA) Transactions on Signal and Information Processing and IEEE MultiMedia Magazine, and Director of the review board and R-Letter in IEEE Multimedia Communications Technical Committee. He also served in multiple IEEE international conferences such as TPC Co-Chair in ICME 2021, Industry Innovation Forum Chair in ICIP 2023, and General Co-Chair in MIPR 2024.  He served as VP Industrial Relations and Development in APSIPA 2018--2019. He has been serving as Vice Chair for Conference in IEEE Technical Committee on Multimedia Computing~(TCMC) since 2021. 
\end{IEEEbiography}

\balance

\newpage

\begin{center}
\textbf{\Large Supplemental Material}    
\end{center}

\section{Test Sequence Summary}
\label{sec:test_sequence_summary}
\underline{\textsc{BasketballDrill}}: A team of players take turns practicing rebounding. The camera positioned at the side of the backboard focuses on the players' shooting layups and the ground.
\underline{\textsc{BQMall}}: A video showing the inside of a mall, with the camera panning to the left, capturing people walking in both directions in front of the camera. 
\underline{\textsc{FourPeople}}: 
Four people are sitting behind a table. Someone at the side of the table takes some flyers from a stand and hands them out to each person.
\underline{\textsc{KristenAndSara}}: 
Two women have a conversation in front of their company poster, and one briefly points to it.
\underline{\textsc{PartyScene}}: The video shows a Christmas party scene, with the camera continuously zooming in. The focus is on a little girl blowing bubbles. At the start of the video, the camera also captures two other children circling a tree. 
\underline{\textsc{Vidyo}}: Three people chat at a table in a conference room; one drinks water and uses a keyboard and a mouse, while a woman walks past the open door.

\section{Determining Search Range for Motion Estimation}
\label{app:search_range}
We determined for each test sequence the search range for the full-search motion estimation algorithm based mainly on the statistics of reference motion vectors. 
We also considered computational complexity when determining the search range.
We used VVC/H.266 to compress each test sequence and obtained motion vectors for reference. We mixed the motion vectors from both the horizontal and vertical directions, calculated such statistics as minimum, maximum, 1-percentile, and 99-percentile, and listed them in TABLE~\ref{tab:search_range}.
Even though the ranges of motion vectors are quite large, e.g., 68 pixels for \textsc{BasketballDrill}, we observed that the ranges were inflated by a few long vectors.
Using the inflated full range of motion vectors as the search range is not efficient, and the quadratic time complexity of full search in the motion vector search range makes it even worse. 
We found that ignoring the longest 2\% of motion vectors can significantly reduce the required motion vector search range. 
We therefore determined, as listed in the last column of TABLE~\ref{tab:search_range}, our search range for most sequences based on the 1-percentile and 99-percentile. One exception is the smaller search range for \textsc{BasketballDrill} to limit the running time of experiments within a reasonable duration.
\begin{table}[!b]
\renewcommand{\arraystretch}{1.3}
\caption{Sequence Statistics and Search Ranges Used.}
\vspace{-2mm}
\label{tab:search_range}
\centering
\scalebox{1.0}{
\begin{tabular}{llcc|c}
\toprule
\multirow{2}{*}{\textbf{Sequence}} & \multicolumn{3}{c|}{\textbf{Video Sequence Statistics}} & \textbf{Search Range}
\\ 
  & \textbf{MV Range} & \textbf{1\%ile} & \textbf{99\%tile} & \textbf{Used}
\\ \hline
BasketballDrill & $[-40, 68]$ & $-26$ & $18$ & $12$
\\ \hline
BQMall & $[-37, 32]$ & $-8$ & $6$ & $10$
\\ \hline
FourPeople & $[-12, 7]$ & $-3$ & $6$ & $7$
\\ \hline
Kristen\&Sara & $[-5, 4]$ & $-2$ & $1$ & $3$
\\ \hline
PartyScene & $[-56, 63]$ & $-8$ & $7$ & $10$
\\ \hline
Vidyo & $[-8, 17]$ & $-2$ & $6$ & $7$
\\ \bottomrule
\end{tabular}
}
\end{table}

\balance

\end{document}